\documentclass[a4paper,11pt]{article}

\usepackage{natbib}
\usepackage{eqnarray,amsmath,amsthm}
\usepackage{amssymb}
\usepackage{graphicx}
\usepackage{epstopdf}
\usepackage{authblk}
\usepackage{mathtools}

\usepackage[ruled,linesnumbered]{algorithm2e}
\SetArgSty{textnormal}
\usepackage{subfigure}
\usepackage{bm}
\usepackage[margin=3cm]{geometry}
\usepackage{xargs}                      
\usepackage[table,pdftex,dvipsnames]{xcolor}  
\usepackage[colorinlistoftodos,prependcaption]{todonotes}

\usepackage[hang,flushmargin]{footmisc}
\makeatletter
\newcommand{\algorithmfootnote}[2][\footnotesize]{%
	\let\old@algocf@finish\@algocf@finish
	\def\@algocf@finish{\old@algocf@finish
		\leavevmode\rlap{\begin{minipage}{\linewidth}
				#1#2
		\end{minipage}}%
	}%
}
\makeatother

\newcommandx{\change}[2][1=]{\todo[linecolor=blue,backgroundcolor=blue!25,bordercolor=blue,#1]{#2}}

\definecolor{best}{RGB}{32,178,170}
\newcommand{\bes}{\cellcolor{best}}

\providecommand{\keywords}[1]{\textbf{\textit{Keywords---}} #1}

\newcommand{\vect}[1]{\boldsymbol{#1}}

\title{\bf Particle Methods for Stochastic Differential Equation Mixed Effects Models}
\author[$\star$,$\ddagger$]{Imke Botha}
\author[$\dagger$,$\ddagger$]{Robert Kohn}
\author[$\star$,$\ddagger$]{Christopher Drovandi}

\affil[$\star$]{School of Mathematical Sciences, Queensland University of Technology}
\affil[$\dagger$]{School of Economics, University of New South Wales}
\affil[$\ddagger$]{Australian Research Council Centre of Excellence for Mathematical \& Statistical	Frontiers (ACEMS)}

\begin{document}
	\setlength{\parindent}{0pc}
	\setlength{\parskip}{1ex}
	
	\maketitle

	\begin{abstract}
		Parameter inference for stochastic differential equation mixed effects models (SDEMEMs) is a challenging problem. Analytical solutions for these models are rarely available, which means that the likelihood is also intractable. In this case, exact inference is possible using the pseudo-marginal method, where the intractable likelihood is replaced by its nonnegative unbiased estimate. A useful application of this idea is particle MCMC, which uses a particle filter estimate of the likelihood.  While the exact posterior is targeted by these methods, a naive implementation for SDEMEMs can be highly inefficient. We develop three extensions to the naive approach which exploits specific aspects of SDEMEMs and other advances such as correlated pseudo-marginal methods. We compare these methods on real and simulated data from a tumour xenography study on mice. 	
	\end{abstract}

	
	\newpage
	\section{Introduction} \label{sec:intro}
	Stochastic differential equations (SDEs) may be defined as ordinary differential equations (ODEs) with one or more stochastic components. SDEs allow for random variations around the mean dynamics specified by the ODE. These models can be used to capture inherent randomness in the system of interest. For repeated measures data, random effects can be used to account for between-subject variability. Assuming measurement error leads to a state-space SDE mixed effects model (SDEMEM). 
	
	SDEMEMs are emerging as a useful class of models for biomedical and pharmacokinetic/pharmacodynamic data \citep{donnet2010bayesian, donnet2013review, leander2015mixed}. They have also been applied in psychology \citep{oravecz2011hierarchical} and spatio-temporal modelling \citep{duan2009modeling}. Statistical inference for these models is generally difficult however. In most cases, the SDE does not have an explicit or analytical solution (transition density), making the likelihood intractable. Including random effects adds further complexity.
	
	Parameter inference for SDEMEMs has largely focussed on maximum likelihood estimation; e.g.\ \citet{picchini2010stochastic}, \citet{picchini2011practical}, \citet{delattre2013maximum} and \citet{donnet2013review, donnet2013pmcmc}. There are few Bayesian inference methods; \citet{donnet2010bayesian} propose a Gibbs sampler coupled with an Euler-Maruyama discretization of the intractable transition density. This approach targets an approximation to the posterior, whose error can be controlled for some models. \citet{whitaker2017bayesian} take a data augmentation approach based on a diffusion bridge, which allows for non-linear dynamics between observed time points. \citet{picchini2019bayesian} compare results from a particle MCMC algorithm \citep{andrieu2009pseudo, andrieu2010particle} and a Bayesian synthetic likelihood approach \citep{wood2010statistical, price2018bayesian}. They apply both methods to an SDE with known solution, and suggest an Euler-Maruyama approximation if the solution is unavailable. 
	
	It is unlikely however that any one approach to estimating SDEMEMs will be optimal for all applications. Performance will depend on the complexity of the underlying SDE, the number of parameters, the number of observations for each subject, as well as the complexity of the random effects. It has been our experience that methods that work well on simple examples can often fail badly on more complex ones. This motivates our article to focus on significant extensions to the pseudo-marginal approach of \citet{picchini2019bayesian} for SDEMEMs.  Pseudo-marginal methods can overcome some limitations of data augmentation approaches because they integrate out the latent states \citep{stramer2011data, gunawan2018copula}. This is especially useful when there is substantial correlation between the latent variables and the parameters of interest. Our article develops a suite of new and efficient Bayesian methods for SDEMEMs by taking advantage of advances in particle methods that can exploit specific aspects of SDEMEMs. As a by-product, we compare the performance of a collection of pseudo-marginal methods for our models of interest. The results of this comparison will be of interest to the wider computational Bayesian community. We compare these methods on a model adapted from one used by \citet{picchini2019bayesian} to model the growth of tumour volumes in mice. 

	The rest of the paper is organised as follows. Sections \ref{sec:bgSDEMEM} and \ref{sec:bgPMM} provide the necessary background on state space models, stochastic differential equations, particle filters and particle MCMC methods. Section \ref{sec:methods} proposes three potential particle methods for SDEMEMs. Sections \ref{sec:ex}-\ref{sec:MCMC} compares these methods with the approach of \citet{picchini2019bayesian} on simulated and real data from a tumor xenography study on mice. Section \ref{sec:disc} concludes with a discussion of the results and possible future work. Code for our methods is available at https://github.com/imkebotha/particle-mcmc-sdemem.
	
	\section{Stochastic Differential Equation Mixed Effects Models} \label{sec:bgSDEMEM}
	We denote random variables by capital letters and their realisations by lowercase letters; $\mathbb{N}$ is the set of positive integers. We use $\sim$ to denote both the distribution and density of a random variable, with the meaning made clear through its context.
	
	\subsection{State Space Models} \label{sec:SSM}
	State space models (SSMs) consist of two processes: a Markov process $\{X_t\}_{t\ge 0}\subset \mathcal{X}^\mathbb{N}$, where $X_t$ is usually only partially observed and is often viewed as a latent process, and an observed process $\{Y_t\}_{t\ge 0}\subset \mathcal{Y}^\mathbb{N}$. The $\mathcal{X}$ and $\mathcal{Y}$ spaces are usually subsets of Euclidean space $\mathbb{R}$ and are often $\mathbb{R}$ itself. To obtain a SSM, we assume that $\{(x_t, y_t);t\ge 0\}$ is Markov with model parameters $\vect\theta$, so that 
	\begin{align*}
	P(x_t, y_t \mid \vect{x}_{0:t-1}, \vect{y}_{0:t-1}, \vect\theta) 
	&= P(x_t, y_t\mid x_{t-1}, y_{t-1}, \vect\theta) \\  
	&= P(y_t\mid x_t, x_{t-1}, y_{t-1}, \vect\theta)
	P(x_t\mid x_{t-1}, y_{t-1}, \vect\theta).
	\end{align*}
	We simplify further and assume that
	\begin{align*}
	P(y_t\mid x_t, x_{t-1}, y_{t-1}, \vect\theta) = g(y_t\mid x_t, \vect\theta), P(x_t\mid x_{t-1}, y_{t-1}, \vect\theta) = f(x_t\mid x_{t-1}, \vect\theta),
	\end{align*}
	where $g(y_t\mid x_t, \vect\theta)$ is the observation density and $f(x_t\mid x_{t-1}, \vect\theta)$ the transition density. Since $\vect\theta$ is unknown, it is assigned a prior $\pi(\vect\theta)$. The unnormalized posterior density of the latent states and model parameters is 
	\begin{align}
	P(\vect{x}_{0:T-1},\vect\theta \mid \vect{y}_{0:T-1}) \propto  
	P(\vect{y}_{0:T-1} \mid \vect{x}_{0:T-1},\vect\theta) P(\vect{x}_{0:T-1} \mid \vect\theta) \pi(\vect\theta),
	 \label{eq:jointpost}
	\end{align}
	where
	\begin{align*}
	P(\vect{y}_{0:T-1} \mid \vect{x}_{0:T-1}, \vect\theta) = &\prod_{t=0}^{T-1}{g(y_t \mid x_t, \vect\theta)} \\
	P(\vect{x}_{0:T-1} \mid \vect\theta) = \mu(x_0 \mid \vect\theta) &\prod_{t=1}^{T-1}{f(x_t\mid x_{t-1},\vect\theta)}.
	\end{align*}
	To obtain parameter inference for $\vect\theta$, we need to consider the marginal posterior,
	\begin{align*}
	P(\vect\theta \mid \vect{y}_{0:T-1} )\propto \pi(\vect\theta)P(\vect{y}_{0:T-1} \mid \vect\theta), 
	\end{align*}
	with likelihood
	\begin{align}
	P(\vect{y}_{0:T-1} \mid \vect\theta)=\int_{X}{P(\vect{y}_{0:T-1} \mid \vect{x}_{0:T-1}, \vect\theta)P(\vect{x}_{0:T-1}\mid \vect\theta)}d\vect{x}_{0:T-1},
	\label{eq:like}
	\end{align}
	especially if high correlation exists between $\vect\theta$ and $\vect{x}_{0:T-1}$. However, this integral is usually intractable. For some models, inference may also be complicated by an intractable transition density, e.g.\ the SDEs in Section \ref{sec:SDEMEM}. While approximate methods can be used in this case, exact inference is still feasible if it is possible to simulate from the transition density (see Section \ref{sec:PM}).
		
	\subsection{Stochastic Differential Equation Mixed Effects Models} \label{sec:SDEMEM}
	It is possible to construct a stochastic differential equation (SDE) from an ordinary differential equation by adding noise or replacing one of the terms in the model by a stochastic process. For simplicity, we decribe a one-dimensional SDE, but it is straightforward to extend the methods introduced in Section \ref{sec:methods} to higher dimensions. Given an It\^{o} process $\{X_t\}_{t\ge 0}$ \citep{oksendal2013stochastic}, the general form for a 1-dimensional continuous SDE is
	\begin{align*}
	dX_t=\mu(X_t,\vect{\phi_X},t)dt+\sqrt{v}(X_t,\vect{\phi_X},t)dB_t,\quad X_0=X_0(\vect{\phi_X})
	\end{align*}
	where $\mu(\cdot)$ is the drift, $\sqrt{v}(\cdot)$ is the diffusion, $\vect{\phi_X}$ are the fixed model parameters for the SDE and $\{B_t\}_{t \ge 0}$ is a standard Brownian motion process. This model can be extended by allowing some of the parameters to vary between the $m=1,\ldots,M$ individuals. In this case we have $\{X_{m,t}\}_{t\ge 0}$ for $m=1,\ldots,M$ instead of $X_t$. In this more general setting, let $\vect{\phi_X}$ be the vector of fixed common parameters of the SDE,  and $\vect{\eta}_m$ the vector of subject specific parameters (random effects), where $\vect{\eta}_m \sim P(\vect{\phi_{\eta}})$. Then, the SDEMEM is given by,
	\begin{align}
	dX_{m,t} = \mu(X_{m,t},\vect{\phi_X},\vect{\eta}_m)dt + \sqrt{v}(X_{m,t},\vect{\phi_X},\vect{\eta}_m)dB_{m,t}, \quad X_{m0}=X_{m0}(\vect{\phi_X},\vect{\eta}_m). 
	\label{eq:SDEMEM}
	\end{align}
	The solution to \eqref{eq:SDEMEM} gives the transition density of the states. If an analytical solution for the transition density is unavailable, numerical methods can be used; some of these are discussed in Section \ref{sec:SDEsim}.  
	
	Equation \eqref{eq:SDEMEM} leads a state-space model as defined in Section \ref{sec:SSM}. Let $y_{m,t}\in\{Y_{m,t}\}$ denote a noisy observation for individual $m, m=1,\ldots,M$ at time $\xi_{m,t}, t = 0,\ldots,T_m-1$, where $T_m$ is the number of observations for individual $m$. To simplify notation, we assume that observations are taken at the same time points for all individuals, i.e.\ $\xi_{t}, t=0,\ldots,T_m-1$, but this restriction is not required for our methods. We assume that the observations are given by 
	\begin{align}
	y_{m,t}\mid  x_{m,t},\sigma^2 \sim \mathcal{N}(y_{m,t} ; x_{m,t},\sigma^2).
	\label{eq:obs_density}
	\end{align}
	Let $\vect\theta=(\sigma,\vect{\phi_X},\vect{\phi_\eta})$ be the vector of all unkown parameters in the model, $\vect{y}_{m} = \vect{y}_{m,0:T_m-1}$ and $\vect{x}_{m} = \vect{x}_{m,0:T_m-1}$. The posterior of $\vect\theta,\vect{\eta}_{1:M}$ can be expressed as 
	\begin{align*}
	P(\vect\theta,\vect{\eta}_{1:M}, \vect{x}_{1:M} \mid \vect{y}_{1:M}) \propto P(\vect\theta) \prod_{m=1}^{M}{P(\vect{y}_m \mid \vect{x}_m,\vect\theta)P(\vect{x}_m \mid \vect\theta,\vect{\eta}_m)P(\vect{\eta}_m \mid \vect\theta).}
	\end{align*}
	
	We will use the following running example throughout the paper to illustrate some of the concepts and methods.
	
	\textbf{Example: SDEMEM with constant drift and diffusion.} Consider the SDEMEM 
	\begin{align}
	&X_{m,t} = \beta_m dt + \gamma dB_t, \quad X_{m, 0} = x_0, 
	\label{eq:simpleSDEMEM} \\
	&\log(\beta_m) \sim \mathcal{N}\left(\log(\beta_m) ; \mu_{\beta}, \sigma_{\beta}^2\right), \nonumber
	\end{align}
	with random effects $\eta_m = \log(\beta_{m})$, unknown static model parameters $\vect{\phi_X} = \{\gamma, x_0\}$ and random effect hyperparameters $\vect{\phi_{\eta}} = \{\mu_{\beta}, \sigma_{\beta}\}$. The exact transition density of this model can be obtained by solving \eqref{eq:simpleSDEMEM}, 
	\begin{align*}
	f(x_{m,t}\mid x_{m,t-1}, \beta_m, \gamma, x_0) = \mathcal{N}(x_{m,t}; x_{m,t-1} + \beta_m, \gamma^2).
	\end{align*}
	If a Gaussian observation density is assumed, the full model is given by
	\begin{align}
	\begin{cases}
		g(y_{m,t}\mid x_{m,t}, \vect\theta) = \mathcal{N}(y_{m,t} ; x_{m,t}, \sigma^2) \\
		f(x_{m, t} \mid x_{m,t-1}, \eta_m, \vect\theta) = \mathcal{N}(x_{m,t} ; x_{m,t-1} + \beta_m, \gamma^2) \\
		P(\eta_m\mid\vect\theta) = \mathcal{N}(\eta_m ; \mu_{\beta}, \sigma_{\beta}^2)
	\end{cases}
	\label{eq:simpleEx}
	\end{align}
	where $\vect\theta = \{\sigma, \gamma, x_0, \mu_{\beta}, \sigma_{\beta}\}$.

	\subsection{SDE Simulation} \label{sec:SDEsim}
	Consider the SDEMEM for a single individual 
	\begin{align*}
	dX_t=\mu(X_t,\vect{\phi_X},\vect\eta)dt + \sigma(X_t,\vect{\phi_X},\vect\eta)dB_t,\quad X_0=X_0(\vect{\phi_X},\vect\eta).
	\end{align*}
	
	If the SDE cannot be solved analytically, then it is necessary to use approximate methods. This section describes two common approaches for approximate simulation of SDEs.
	
	\subsubsection{Euler-Maruyama} \label{sec:EM}
	The Euler-Maruyama discretization (EMD) is the simplest method for simulating approximately from an SDE. Given a process $\{X_t\}_{t\ge0}$, the time interval $[0,J]$ is split into $D$ subintervals,
	\begin{align*}
	0=\tau_0<\tau_1<\cdots<\tau_k<\tau_{k+1}<\cdots<\tau_D=J, \quad \Delta\tau=\frac{J}{D}.
	\end{align*}
	Assuming that the drift and diffusion coefficients are locally constant, 
	\begin{align*}
	\mu(X_{\tau_k},\vect{\phi_X},\vect\eta)&=\mu_k \\
	\sqrt{v}(X_{\tau_k},\vect{\phi_X},\vect\eta)&=\sqrt{v_k},
	\end{align*}
	the EMD simulates over each subinterval as follows
	\begin{align*}
	X_{\tau_{k+1}}&=X_{\tau_k} + \mu_k\Delta\tau + \sqrt{v_k}\Delta B_{\tau_k} \\
	\Delta B_{\tau_k}&= B_{\tau_{k+1}}-B_{\tau_k}.
	\end{align*}	
	Since $\Delta B_{\tau_k} \sim \mathcal{N}(\Delta B_{\tau_k} ; 0,\Delta\tau)$ by definition, the path is simulated through a recursive application of 
	\begin{align*}
	x_{\tau_{k+1}} \mid x_{\tau_k} \sim \mathcal{N}(x_{\tau_{k+1}}; x_{\tau_k}+\mu_k\Delta\tau, v_k\Delta\tau).
	\end{align*}
	Thus, the joint density of this approximation is
	\begin{align*}
	q(x_{\tau_1:J} \mid x_0, \phi_X, \eta) \propto \prod_{k=0}^{D-1}{\mathcal{N}(x_{\tau_{k+1}};x_{\tau_k}+\mu_k\Delta\tau,v_k\Delta\tau)};
	\end{align*}
	we note that for an SDE with constant drift and diffusion, the EMD gives the exact solution. 
	
	\textbf{Example: SDEMEM with constant drift and diffusion.} For the SDEMEM in Equation \eqref{eq:simpleSDEMEM}, $\mu_k = \beta_m$ and $v_k = \gamma^2$. The EMD of this model is 
	\begin{align*}
	x_{m,\tau_{k+1}} \mid x_{m,\tau_k} &\sim \mathcal{N}(x_{m,\tau_{k+1}} ; x_{m,\tau_k}+\beta_m\Delta\tau, \gamma^2\Delta\tau), \\
	x_{m,\tau_{k+1}} \mid x_{m,\tau_k} &\sim \mathcal{N}(x_{m,\tau_{k+1}} ; x_{m,\tau_k}+\beta_m, \gamma^2), \quad \Delta\tau = 1,
	\end{align*}
	which in this case is the exact transition density. 
	
	\subsubsection{Diffusion Bridges} \label{sec:DB}
	Simulating from the (approximate) transition density may not perform well in pseudo-marginal methods if any particular observations are highly informative or there is little observation noise. More efficient estimates of the likelihood of SSMs can be achieved if the proposal for $x_t$ can be directed towards $y_t$. One option to do this is to use a diffusion bridge. 
	
	The modified diffusion bridge (MDB) of \citet{durham2002numerical} \citep[see also][]{golightly2008bayesian} is derived by approximating the joint distribution of $X_{\tau_{k+1}},Y_J \mid x_{\tau_k}$ using multivariate normal theory, and then conditioning on $Y_J=y_J$. The density $X_{\tau_{k+1}},Y_J \mid x_{\tau_k}$ is obtained from the observation density \eqref{eq:obs_density} and the EMD of $X_{\tau_{k+1}} \mid x_{\tau_k}$. See Appendix 1 of \citet{golightly2008bayesian} for a more detailed derivation. The MDB gives a bridge proposal of the form
	\begin{align*}
	x_{\tau_{k+1}} \mid x_{\tau_k}, y_J \sim \mathcal{N}\{x_{\tau_{k+1}}; x_{\tau_k}+\mu_{\textrm{MDB}}(x_{\tau_k}, y_J)\Delta\tau, \Psi_{\textrm{MDB}}(x_{\tau_k})\Delta\tau\}
	\end{align*}
	where $\Delta_k=J-\tau_k$,
	\begin{align*}
	\mu_{\textrm{MDB}}(x_{\tau_k}, y_J)&=\mu_k+\frac{v_k(y_J-(x_{\tau_k}+\mu_k\Delta_k))}{v_k\Delta_k+\sigma^2} &= \frac{\mu_k\sigma^2 + v_k(y_J-x_{\tau_k})}{v_k\Delta_k+\sigma^2} \\
	\Psi_{\textrm{MDB}}(x_{\tau_k})&=v_k-\frac{v_k^2\Delta\tau}{v_k\Delta_k+\sigma^2} &= \frac{v_k\sigma^2 + v_k^2(\Delta_k - \Delta\tau)}{v_k\Delta_k+\sigma^2}.
	\end{align*}
	
	\citet{whitaker2017improved} notes that the modified diffusion bridge can perform poorly when the drift coefficient is not approximately constant. To overcome this problem, they propose to partition the SDE into a deterministic process and a residual stochastic process, such that the latter has constant drift. Rewriting the model in terms of these processes gives
	\begin{align}
	X_t&=\zeta_t + R_t,\quad &\zeta_t,t\ge 0 \nonumber\\
	d\zeta_t&=f(\zeta_t)dt, \quad &\zeta_0=x_0 \nonumber\\
	dR_t&=\{\mu(X_{t},\vect{\phi_X},\vect\eta)-f(\zeta_t)\}dt + \sigma(X_{t},\vect{\phi_X},\vect\eta)dB_t,\quad &R_0=0. 
	\label{eq:residSDE}
	\end{align}
	The idea is to choose $\zeta_t$ and $f(\cdot)$ such that the drift of \eqref{eq:residSDE} is approximately constant. The simplest solution \citep{whitaker2017improved} is to set $\zeta_t=\eta_t$ and $f(\cdot)=\mu(\cdot)$ as
	\begin{align*}
	X_t&=\eta_t+R_t, \quad &\eta_t,t\ge 0 \\
	d\eta_t&=\mu(\eta_t,\vect{\phi_X},\vect\eta)dt, \quad &\eta_0=x_0 \\
	dR_t&=\{\mu(X_{t},\vect{\phi_X},\vect\eta)-\mu(\eta_t,\vect{\phi_X},\vect\eta)\}dt + \sigma(X_{t},\vect{\phi_X},\vect\eta)dB_t,\quad &R_0=0
	\end{align*}
	noting that $Y_J-\eta_J=R_J+\epsilon_J$. The residual bridge is obtained by constructing the MDB on the residual process $\{R_t\}$. This gives a bridge proposal of the form
	\begin{align*}
	x_{\tau_{k+1}} \mid x_{\tau_k},y_J \sim \mathcal{N}(x_{\tau_{k+1}}; x_{\tau_k}+\mu_{\textrm{RB}}(x_{\tau_k},y_J)\Delta\tau,\Psi_{\textrm{RB}}(x_{\tau_k,y_J})\Delta\tau),
	\end{align*}
	where 
	\begin{align*}
	\Psi_{\textrm{RB}}(x_{\tau_k},y_J)&=\Psi_{\textrm{MDB}}(x_{\tau_k},y_J), \delta_k^\eta = \frac{\eta_{\tau_{k+1}}-\eta_{\tau_k}}{\Delta\tau} \quad \textrm{and}\\
	\mu_{\textrm{RB}}(x_{\tau_k},y_J)&=\mu_k+\frac{v_k(y_J-(\eta_J+r_{\tau_k}+(\mu_k-\delta_k^\eta)\Delta k))}{v_k\Delta_k+\sigma^2}. 
	\end{align*}

	\section{Particle MCMC} \label{sec:bgPMM}
	
	\subsection{Particle Filters} \label{sec:PF}	
	Exact state estimation of SSMs using the Kalman filter is only possible when they are Gaussian or conditionally Gaussian. In the case of non-linear, non-Gaussian SSMs, a particle filter can be used for simulation consistent estimation \citep{gorden1993BPF,carpenter1999nonlinear,doucet2000, del2006sequential,doucet2009tutorial}. 

	Particle filters are used to traverse through a sequence of intermediary distributions towards some target distribution. We describe the generic particle filter of \citet{doucet2009tutorial} (see Algorithm \ref{alg:genPF}), with filtering distribution of the form 	
	\begin{align}
	\pi_t(\vect{x}_{1:t} \mid \vect{y}_{1:t}, \vect\theta) &\propto \pi_t(\vect{x}_{1} \mid \vect{y}_{1}, \vect\theta) 
	\prod_{j=2}^{t}{\pi_t(x_{j} \mid y_{j}, x_{j-1}, \vect\theta)}, \quad t=1,\ldots,T \nonumber \\
	&= g(y_1|x_1,\vect\theta) f(x_1 \mid \vect\theta) \prod_{j=2}^{t} g(y_j|x_j,\vect\theta) f(x_j \mid x_{j-1}, \vect\theta).
	\label{eq:PFpi}
	\end{align}
	A combination of move, reweight and resample steps are used to transition through this sequence. The move step generates values for $x_t$ from some proposal distribution $q(x_t\mid y_t, x_{t-1},\vect\theta)$. Once moved, the $N$ particles are re-weighted according to, 
	\begin{align*}
	w_t^n = W_{t-1}^n\frac{\pi_t(x_t\mid y_t, x_{t-1}\vect\theta)}{q(x_t \mid y_t, x_{t-1},\vect\theta)}, \quad W_t^n=\frac{w_t^n}{\sum_{i=1}^{N}{w_t^i}}.
	\end{align*}

	Particles are then resampled with probability $\vect{W}_t^{1:N}$ for the next iteration. This is done to avoid particle impoverishment, where most of the weight is given to few particles. There are several resampling methods that can be used, including multinomial, stratified \citep{kitagawa1996monte}, and more recently, Srinivasan \citep{gerber2019negative}. 
	
	An attractive feature of the particle filter is that an unbiased estimate of the likelihood may be obtained from the unnormalized weights,
	\begin{align*}
	\widehat P(\vect{Y}_{1:T} \mid \vect\theta) = \prod_{t=1}^{T}{\sum_{n=1}^{N}{w_t^{(n)}}}.
	\end{align*}
	
	The bootstrap particle filter (BPF) of \citet{gorden1993BPF} is a special case of the PF with $q(x_t \mid y_t, x_{t-1},\vect\theta) = f(x_t\mid x_{t-1},\vect\theta)$. The calculation of the weights then simplifies to $w_t^n=W_{t-1}^n g(y_t\mid x_t,\vect\theta)$.
	
	\begin{algorithm}[htp]
		\SetKwInOut{Input}{Input}
		
		\Input{data $\vect{y}_{1:T}$, the number of particles $N$, the static parameters $\vect\theta$ and the initial state $x_0$. We use the convention that index $(n)$ means 'for all $n\in \{1,\ldots,N\}$'}
		
		Initialise $x_1^{(n)}=x_0, \ W_1^{(n)}=\frac{1}{N}, \  w_1^{(n)}=W_1^{(n)}P(y_1|x_1^{(n)},\vect\theta), \ Z=\sum_{n=1}^{N}{w_1^{(n)}}$ \\
		
		\For{$t=2$ to $T$}{
			Resample (with replacement) $N$ particles from $\vect{x}_{t-1}^{1:N}$ according to $\vect{W}_{t-1}^{1:N}$ \\
			
			Move the particles, $x_t^{(n)}\sim q(x_t^{(n)}\mid x_{t-1}^{(n)}, \vect\theta)$ \\
			Calculate weights $w_t^{(n)} = W_{t-1}^{(n)}\frac{\pi_t(x_t^{(n)}\mid y_t, x^{(n)}_{t-1}\vect\theta)}{Nq(x_t^{(n)}\mid y_t,  x_{t-1}^{(n)},\vect\theta)}$ \\
			
			Normalize weights $W_t^{(n)}=\frac{w_t^{(n)}}{\sum_{i=1}^{N}{w_t^{(i)}}}$ \\ 
			
			Update likelihood estimate $Z = Z \times \sum_{n=1}^{N}{w_t^{(n)}}$ \\
			
		}
		\caption{The generic particle filter of \citet{doucet2009tutorial}.}
		\label{alg:genPF}
	\end{algorithm}

	\subsection{Pseudo-Marginal MCMC} \label{sec:PM}
	The pseudo-marginal approach of \citet{andrieu2009pseudo} allows for exact inference for models with intractable likelihoods. In this approach, the intractable likelihood $P(\vect{y}_{1:T}\mid\vect\theta)$ is replaced with a nonnegative unbiased estimate of the form $\widehat{P}(\vect{y}_{1:T}\mid\vect\theta)$, which we write as $P(\vect{y}_{1:T}\mid\vect\theta, \vect u)$, where $\vect u\sim P(\vect u)$ are the auxiliary variables used to estimate the likelihood. Since this estimate is unbiased, $\mathbb{E}_{P(\vect{u})}(P(\vect{y}_{1:T}\mid\vect\theta, \vect u)) = P(\vect {y}_{1:T}\mid\vect\theta)$. Pseudo-marginal MCMC can therefore be defined as standard MCMC on an augmented space, i.e.\ the space of $\vect\theta$ augmented with $\vect u$. The chain targets $P(\vect\theta, \vect u\mid \vect{y}_{1:T})$ which has the posterior $P(\vect\theta\mid \vect{y}_{1:T})$ as marginal distribution, as
	\begin{align*}
	\int{P(\vect\theta, \vect u\mid \vect {y}_{1:T})}d\vect u &= \int{\frac{P(\vect {y}_{1:T}\mid \vect\theta, \vect u)P(\vect\theta)P(\vect u)}{P(\vect {y}_{1:T})}}d\vect u \\
	&= \frac{P(\vect\theta)}{P(\vect {y}_{1:T})} \int{P(\vect {y}_{1:T}\mid \vect\theta, \vect u)P(\vect u)}d\vect u \\
	&= \frac{P(\vect\theta)P(\vect {y}_{1:T}\mid\vect\theta)}{P(\vect {y}_{1:T})} \\
	&= P(\vect\theta\mid\vect {y}_{1:T}).
	\end{align*}
	
	The next sections describes the particle marginal Metropolis-Hastings (PMMH) and particle Gibbs (PG) algorithms proposed by \citet{andrieu2010particle}.

	\subsubsection{Particle Marginal Metropolis-Hastings} \label{sec:PMMH}
	The PMMH method is a Metropolis-Hastings algorithm where the intractable likelihood is replaced by its unbiased estimate (see Section \ref{sec:methods} for its use in our particle filter application). As discussed in Section \ref{sec:PM}, the resulting chain targets the joint density $P(\vect\theta,\vect u \mid \vect {y}_{1:T})$, where $\vect u$ is the vector of random numbers used in the particle filter. Unbiasedness implies the posterior of interest $P(\vect \theta \mid \vect {y}_{1:T})$ is obtained through marginalisation.

	\begin{algorithm}[htp]

		Initialise $\vect {\theta}^{(0)}$ \\
		
		Run Algorithm \ref{alg:genPF} to obtain an unbiased estimate of $\widehat P(\vect {y}_{1:T} \mid \vect {\theta}^{(0)})$ \\
		
		\For{$i=1$ to $I-1$} {
			Sample $\vect{\theta^*} \sim q(\cdot\mid\vect{\theta}^{(i-1)})$ \\
			
			Run Algorithm \ref{alg:genPF} to obtain an unbiased estimate of $\widehat P(\vect {y}_{1:T}\mid\vect{\theta^*})$ \\
			
			Calculate the Metropolis-Hastings ratio 
			\begin{eqnarray*}
			\textrm{MHR}=\frac{\widehat P(\vect {y}_{1:T}\mid\vect{\theta^*})P(\vect{\theta^*})}{\widehat P(\vect {y}_{1:T}\mid\vect{\theta}^{(i-1)})P(\vect{\theta}^{(i-1)})}\frac{q(\vect{\theta}^{(i-1)}\mid\vect{\theta^*})}{q(\vect{\theta^*}\mid\vect{\theta}^{(i-1)})}
			\end{eqnarray*}
			
			Draw $u\sim \mathcal{U}(0,1)$ \\
			
			\eIf{$u<\text{MHR}$}{
				Set $\vect{\theta}^{(i)}=\vect{\theta^*}$ \\
			}{
			Set $\vect{\theta}^{(i)}=\vect{\theta}^{(i-1)}$ \\
			} 
			
		}
		\caption{Particle marginal Metropolis-Hastings.}
		\label{alg:pmmh}
	\end{algorithm}

	A drawback of the PMMH algorithm is that it can be difficult to find good proposals. Another drawback is the chain's tendency to get stuck whenever the likelihood is greatly overestimated for a particular value of $\vect\theta$, i.e.\ if $\widehat P(\vect {y}_{1:T}\mid \vect {\theta})$ is greatly overestimated, then the acceptance probability for $\vect\theta^*$ will be very small unless $\widehat P(\vect {y}_{1:T}\mid \vect {\theta^*})$ is also overestimated. This can be mitigated by decreasing the variance of the log of the ratio of the likelihood estimates
	\begin{align}
	R = \log{\left(\frac{\widehat P(\vect {y}_{1:T}\mid \vect {\theta^*})}{\widehat P(\vect {y}_{1:T}\mid \vect \theta)}\right)}. 
	\label{eq:ll_ratio}
	\end{align}
	A common strategy to do this is to increase the number of particles used in the particle filter. \citet{sherlock2015efficiency}, \citet{pitt2012some} and \citet{doucet2015efficient} showed that optimal performance (for random walk proposals) is gained when $N$ is chosen such that the standard deviation of the estimated log-likelihood is between $1$ and $2$. An alternative approach is the correlated pseudo-marginal (CPM) method of \citet{deligiannidis2018correlated} (see also \citet{dahlin2015accelerating}). \citet{tran2016block} introduced a variation of the CPM method called the block pseudo-marginal (BPM) approach. The BPM has a natural application to SDEMEMs, which is discussed further in Section \ref{sec:CPM}. 

	\subsubsection{Correlated Pseudo-Marginal} \label{sec:CPM}
	
	Recall that the chain targets the density $p(\vect\theta,\vect u|\vect {y}_{1:T})$. At iterations $i$ and $i+1$, the estimates returned are proportional to $\widehat P(\vect{\theta}^{(i)},\vect {u}^{(i)} \mid \vect {y}_{1:T})$ and $\widehat P(\vect{\theta}^{(i+1)},\vect {u}^{(i+1)}\mid \vect {y}_{1:T})$. \citet{deligiannidis2018correlated} show that the mixing of the chain can be improved by correlating $\widehat P(\vect{\theta}^{(i)},\vect {u}^{(i)}\mid \vect {y}_{1:T})$ and $\widehat P(\vect{\theta}^{(i+1)},\vect {u}^{(i+1)}\mid \vect {y}_{1:T})$. This helps to vastly reduce the variance of \eqref{eq:ll_ratio}, without having to reduce the variance of the individual log-likelihood estimates. 
	
	The CPM approach correlates these estimates by making $\vect {u}^{(i)}$ and $\vect {u}^{(i+1)}$ highly correlated. Assuming the random numbers are normally distributed, \citet{deligiannidis2018correlated} use the Crank-Nicolson (CN) proposal to induce the correlation
	\begin{align*}
	q_{\vect\theta,\vect u}(\{\vect{\theta^*},\vect {u^*}\} \mid \{\vect\theta,\vect u\}) = 
	q_{\vect\theta}(\vect{\theta^*} \mid \vect\theta) q_{\vect{u}}(\vect {u^*}\mid \vect u) \\= 
	q_{\vect\theta}(\vect{\theta^*} \mid \vect\theta) \mathcal{N}(\vect {u^*}; \sqrt{1-\sigma_u^2}u,\quad \sigma_u^2\vect {I_{N_u}}).
	\end{align*}
	If the particle filter depends on non-normal random numbers, transformations to normality are applied. 
	
	In BPM, correlation is induced by updating $\vect u$ in blocks \citep{tran2016block}. In this approach, the vector of random numbers $\vect u$ is divided into $B$ blocks, and a single block is updated at each iteration while the remaining $B-1$ are held constant. The resulting correlation between the logs of the likelihood estimates is approximately $1-1/B$ and is induced much more directly than CPM. No assumption about the form or distribution of $\vect u$ is required. This approach has a natural application to SDEMEMs, as the blocks can be defined using the subjects, i.e.\ each block contains all random numbers needed to estimate the likelihood for one or more subjects.
	
	Relative to standard PMMH, both CPM and BPM are able to tolerate significantly more variance in the log-likelihood estimates, such that less particles are needed for the chain to mix well. The increase in computational efficiency gained from this typically outweighs the overhead associated with storing the vector of random numbers $\vect u$. 
	
	The number of particles $N$ needed for CPM and BPM can be tuned using the log-likelihood ratio \eqref{eq:ll_ratio} \citep{deligiannidis2018correlated}. To minimize the distance between successive log-likelihood estimates, the number of particles $N$ may be chosen such that the variance of \eqref{eq:ll_ratio} is around 1. 
	
	\subsubsection{Conditional Particle Filter} \label{sec:CSMC}
	The particle Gibbs (PG) algorithm of \citet{andrieu2010particle}, requires a variation of the generic PF (Section \ref{sec:PF}) called the conditional particle filter (CPF). The CPF differs from the generic PF by holding a single path $\vect {x}_{1:T}^k$ invariant throughout the iterations. See Algorithm \ref{alg:CSMC} for more details. 
	
	Once a weighted sample is obtained, a new invariant path may be drawn using the backwards sampling method of \citet{whiteley2010discussion} and \citet{lindsten2012use}; see Algorithm \ref{alg:bsampler}. 
	
	\begin{algorithm}[htp]
		\SetKwInOut{Input}{Input}
		
		\Input{data $\vect {y}_{1:T}$, number of particles $N$, initial state $x_0$, static parameters $\vect\theta$, invariant path $\vect {x}_{1:T}^k$ and associated ancestral lineage $\vect {b}_{1:T}^k$. We use the convention that index $(n\ne k)$ means `for all $n\in \{1,\ldots,k-1,k+1,\ldots,N\}$'}
		
		Initialise $x_1^{(n\ne B_1^k)}=x_0, \ W_1^{(n)}=\frac{1}{N}, w_1^{(n)}=W_1^{(n)}P(y_1\mid x_1^{(n)},\vect\theta), Z=\sum_{n=1}^{N}{w_1^{(n)}}$ \\
		
		\For{$t=2$ to $T$} {
			Sample parent indices $A_{t-1}^{(n\ne B_1^k)}\sim \mathcal{F}(\cdot\mid W_{t-1}^{(n)})$ \tcc*[f]{resampling step}
			
			Sample $x_t^{(n\ne B_1^k)}\sim q(\cdot\mid x_{t-1}^{A_{t-1}^{B_1^n}},\vect\theta)$	\tcc*[f]{move step}
			
			Calculate weights $w_t^{(n)}=\frac{\pi_t(x_t^{(n)}\mid y_t, x^{(n)}_{t-1}, \vect\theta)}{Nq(x_t^{(n)}\mid x_{t-1}^{(n)},\vect\theta)}$ \\
			
			Normalize weights $W_t^{(n)}=\frac{w_t^{(n)}}{\sum_{i=1}^{N}{w_t^{i}}}$
		}		
		Run Algorithm \ref{alg:bsampler} to obtain new ancestral lineage $\vect {b}_{1:T}^{k^*}$ \\
		
		Use $\vect b_{1:T}^{k^*}$ to determine new path $\vect {x}_{1:T}^{k^*}$
		\caption{The conditional particle filter. }
		\algorithmfootnote{The matrix $A_{t-1}^n$ gives the parent indices of the particles at time $t-1$. The relationship between the ancestral lineage and the matrix of parent indices is $A_{t-1}^{B_t^k}=B_{t-1}^k$, where $B_T^k=k$.}
		\label{alg:CSMC}
	\end{algorithm}

	\begin{algorithm}[htp]
		\SetKwInOut{Input}{Input}
		\SetKwInOut{Output}{Output}
		\Input{$\vect {w}_{1:T}^{(n)}, \ \vect {W}_T^{(n)}$}
		\Output{a new ancestral lineage $\vect {B}_{1:T}^{k^*}$}
		
		Draw $k^*\sim \mathcal{F}(\cdot\mid W_T)$ \\
		Set $B_{T}^{k^*}=k^*$ \\
		
		\For{$t=T-1$ to $1$} {
			Sample $W_{(t\mid T)}^{(n)}=w_t^{(n)}\frac{f_{\vect\theta}\left(x_{t+1}^{B_{t+1}^{k^*}} \mid x_t^{(n)}\right)}{\sum_{i=1}^{N}{w_t^{(i)}f_{\vect\theta}\left(x_{t+1}^{B_{t+1}^{k^*}} \mid x_t^{(i)}\right)}}$ \\
			
			Draw $B_t^{k^*}\sim\mathcal{F}(\cdot\mid W_{(t\mid T)})$
		}
		\caption{Backward Sampling.}
		\label{alg:bsampler}
	\end{algorithm}

	\subsubsection{Particle Gibbs} \label{sec:PG}
	In PMMH, the particle filter returns an estimate of the likelihood \eqref{eq:like}. In particle Gibbs, the latent states are updated using a conditional particle filter, i.e.\ $\vect {x}_{1:T}$ is approximately sampled from $p(\vect {x^*}_{1:T}\mid \vect {y}_{1:T}, \vect {x}_{1:T}, \vect\theta)$ (see Algorithms \ref{alg:CSMC} and \ref{alg:pg}). The parameters $\vect\theta$ may be updated using Gibbs sampling if the full conditional posterior is available, or a Metropolis-Hastings step if it is not. 
	
	\begin{algorithm}[htp]
		
		Initialise $\vect{\theta}^{(0)}, \vect {x}_{1:T}^{(0)}$ and associated ancestral lineage $\vect {b}_{1:T}^{(0)}$ \\

		\For{$i=1$ to $I-1$} {
			Update $\vect{\theta}^{(i+1)}$ conditional on $\vect{\theta}^{(i)}$ and $\vect {x}_{1:T}^{(i)}$ \\
			
			Run Algorithm \ref{alg:CSMC} to sample $\vect {x}_{1:T}^{(i+1)}$ and $\vect {b}_{1:T}^{(i+1)}$ conditional on $\vect{\theta}^{(i+1)}, \vect {x}_{1:T}^{(i)}$ and $\vect {b}_{1:T}^{(i)}$. \\ 
		}	
		\caption{The particle Gibbs algorithm.}
		\label{alg:pg}
	\end{algorithm}
	Since a new path $\vect {x}_{1:T}$ is simulated at each iteration, PG does not suffer from the same mixing problem as PMMH. As such, it is significantly less sensitive to the number of particles used. PG also has the advantage that more efficient updating schemes for $\vect\theta$ can be used, such as MALA or HMC. While this method has a number of advantages over PMMH, it is not as generally applicable as a closed form transition density is required to update $\vect\theta$.

	\section{Methods} \label{sec:methods}
	We are interested in parameter inference for the state-space SDEMEM described in Section \ref{sec:SDEMEM}. For a single individual $m$, with observations taken at $\xi_t, t=0,\ldots,T_m-1$ and level of discretisation $D$, the sequence of distributions \eqref{eq:PFpi} traversed by the particle filter (see Section \ref{sec:PF}) is
	\begin{align*}
	\pi_t(\vect {x}_{m,0:t}  \mid \vect {y}_{m,0:t}, \vect{\eta}_m, \sigma, \vect{\phi_X}) \propto
	g(y_{m,0}|x_{m,0},\sigma)f(x_{m,0}\mid \vect{\eta}_m, \vect{\phi_X})
	\prod_{k=1}^{D-1}{f(x_{m, k/D}\mid x_{m,(k-1)/D}, \vect{\eta}_m, \vect{\phi_X})} \\
	\prod_{j=1}^{t} \left(g(y_{m,j}|x_{m,j},\sigma)\prod_{k=0}^{D-1}{f(x_{m,j + k/D}\mid x_{m,j + (k-1)/D}, \vect{\eta}_m, \vect{\phi_X})}\right).\\
	\end{align*}
	This particle filter returns an estimate of $P(\vect {y}_{m} \mid \vect {\eta}_m, \sigma, \vect {\phi_X})$. The estimated likehood for all the data $\vect {y}_{1:M}$ is given by
	\begin{align*}
	\widehat P(\vect {y}_{1:M} \mid \vect {\eta}_{1:M}, \sigma, \vect {\phi_X}) = \prod_{m=1}^{M}{\widehat P(\vect {y}_{m} \mid \vect{\eta}_m, \sigma, \vect{\phi_X})}.
	\end{align*}

	\subsection{Individual-Augmentation Pseudo-Marginal} \label{sec:IAPM}
	The first method is Individual-Augmentation Pseudo-Marginal (IAPM), named for the additional auxiliary variables required to estimate the likelihood for each individual. Here, we use the likelihood estimate,
	\begin{align*}
		\widehat P(\vect{{y}}_{m} \mid \vect \theta) & =
		\int{\widehat P(\vect {y}_{m} \mid \vect {\eta}_m, \sigma, \vect {\phi_X}) P(\vect {\eta}_m \mid \vect {\phi_\eta}) g(\vect {\eta}_m \mid \vect \theta)}d\vect{\eta}_m, &\quad \vect\theta = (\sigma, \vect{\phi_X}, \vect{\phi_\eta}) \\
		& \approx \frac{1}{L}{\sum_{l=1}^{L}
			{\frac{\widehat P(\vect {y}_{m} \mid \vect {\eta}_m^{(l)}, \sigma, \vect {\phi_X}) P(\vect {\eta}_m^{(l)} \mid \vect{\phi_\eta})}{g(\vect{\eta}_m^{(l)} \mid \vect\theta)}}}, 
		&\quad \vect{\eta}_m^{(l)}\sim g(\vect{\eta}_m \mid \vect\theta) 
	\end{align*}
	with importance distribution $g(\vect{\eta}_m \mid \vect\theta)$ within a PMMH algorithm (Algorithm \ref{alg:pmmh}). See Algorithms \ref{alg:IAPM} and \ref{alg:iapmIS} for more details.
	
	\begin{algorithm}[htp]
		
			initialise $\vect {\theta}^{(0)}$ \\
			Run Algorithm \ref{alg:iapmIS} to obtain likelihood estimate $\widehat P(\vect {y}_{1:M} \mid \vect{\theta}^{(0)})$ \\
		\For{$i=1$ to $I$} {
			
			Draw $\vect{\theta^*} \sim q(\cdot \mid \vect{\theta}^{(i-1)})$ \\
			
			Run Algorithm \ref{alg:iapmIS} to obtain likelihood estimate $\widehat P(\vect {y}_{1:M} \mid \vect{\theta^*})$ \\
			
			Accept $\vect{\theta^*}$ with probability 
			\begin{align*}
			\alpha = \min\left(1,\frac{\widehat P(\vect {y}_{1:M} \mid \vect{\theta^*}) P(\vect{\theta^*}) q(\vect{\theta}^{(i-1)} \mid \vect{\theta^{*}})}{\widehat P(\vect {y}_{1:M} \mid \vect{\theta}^{(i-1)}) P(\vect{\theta}^{(i-1)}) q(\vect{\theta^*} \mid \vect{\theta}^{(i-1)})} \right)
			\end{align*}
		}	
		\caption{The individual-augmentation pseudo-marginal method.}
		\label{alg:IAPM}
	\end{algorithm}
	
	\begin{algorithm}[htp]
		\For{$m=1$ to $M$} {
			\For{$l=1$ to $L$} {
				Draw $\vect{\eta}_m^{(l)} \sim g(\cdot \mid \vect\theta)$ \\
				
				Run Algorithm \ref{alg:genPF} with $\vect{\eta}_m^{(l)}$ to obtain the likehood estimate $Z_m^{(l)}$ \\
				
				Correct for the importance distribution $Z_m^{(l)} = \frac{Z_m^{(l)}}{g(\vect{\eta}_m^{(l)} \mid \vect\theta)}$
			}	
		
			Calculate $\widehat P(\vect {y}_{m} \mid \vect \theta) = \frac{1}{L}\sum_{i=1}^{L}{Z_m^{(i)}}$ \\
		}
		Calculate $\widehat P(\vect {y}_{1:M} \mid \vect\theta) = \prod_{m=1}^{M}{\widehat P(\vect {y}_{m} \mid \vect \theta)}$
		\caption{Estimating the likelihood for the IAPM algorithm.}
		\label{alg:iapmIS}
	\end{algorithm}
	
	The variability of $\widehat P(\vect {y}_{m} \mid \vect\theta)$ for a given $g(\vect {\eta}_m \mid \vect \theta)$ is controlled by the number of particles $N$, as well as the number of random effects draws $L$. The choice of importance distribution $g(\cdot\mid\vect\theta)$ has an important impact on both of these quantities. A naive choice is $g(\vect{\eta}_m \mid \vect\theta) = P(\vect{\eta}_m \mid \vect\theta)$. While this simplifies the likelihood calculation, it can be very inefficient if $\widehat P(\vect{\eta}_m \mid \vect {y}_{m}, \vect\theta)$ and $P(\vect{\eta}_m \mid \vect\theta)$ are not similar. We propose instead to use a Laplace approximation of a distribution over $\vect{\eta}_m$ that is proportional to
	\begin{align*}
		P(\vect {y}_{m} \mid \vect{{\widehat{x}}}_m, \vect\theta) P(\vect{\eta}_m \mid \vect\theta),
	\end{align*}
	where $\vect{{\widehat{x}}}_m$ is an approximation of $\vect {x}_m$. We present two choices for $\vect{\widehat{x}}_m$. The first uses the solution of the ODE given by the drift of the SDEMEM \eqref{eq:SDEMEM},
	\begin{align*}
		d\widehat{X}_{m,t} = \mu(\widehat{X}_{m,t}, \vect {\phi_X}, \vect{\eta}_m)dt.
	\end{align*}
	The second approximates $\vect {x}_m$ with the mean of the modified diffusion bridge (see Section \ref{sec:DB}), with $\Delta_k = \Delta_t = \tau_{t+1} - \tau_{t}$, such that
	\begin{align*}
	\widehat{x}_{m,t+1} = \widehat{x}_{m,t} + \mu_{\textrm{MDB}}(\widehat{x}_{m,t})\Delta_t
	= \widehat{x}_{m,t} + \frac{\mu_t\sigma^2 + v_t(y_{m,t+1}-\widehat{x}_{m,t})}{v_t \Delta_t+\sigma^2}\Delta_t.
	\end{align*}
	We refer to these importance distributions as Laplace-ODE and Laplace-MDB respectively. 
	
	As a variance reduction technique, randomised quasi-Monte Carlo (RQMC) can be used to draw $\vect{\eta}_m^{(l)}$ (step 2 of Algorithm \ref{alg:iapmIS}). See \citet{L2016randomized} for an overview of RQMC.
	
	A correlated version of IAPM (cIAPM) is possible using block pseudo-marginal as described in Section \ref{sec:CPM}. Here, the vector of random numbers is given by $\vect u = (\vect {u_{\mathrm{RE}}}, \vect {u_{\mathrm{PF}}})$, where $\vect {u_{\mathrm{RE}}}$ and $\vect {u_{\mathrm{PF}}}$ are the random numbers used to draw the random effects and those used in the particle filter respectively. At each iteration of the chain, new random numbers for individual $m, 1\le m \le M$ are proposed, while the rest are held constant. This induces a correlation of approximately $1-\frac{1}{M}$ between successive log-likelihood estimates \citep{tran2016block}. Since BPM makes no assumptions about the distribution of $\vect u$, RQMC is straightforward to use within cIAPM.
	
	\textbf{Example: SDEMEM with constant drift and diffusion.} For the SDEMEM in Equation \eqref{eq:simpleSDEMEM}, the IAPM approximation of $P(\vect {y}_m\mid\vect\theta)$ with importance distribution $g(\vect{\eta}_m \mid \vect\theta)$ is given by
	\begin{align*}
	\frac{1}{L}{\sum_{l=1}^{L}
		{\frac{\widehat P(\vect {y}_{m} \mid \vect {\beta}_m^{(l)}, \sigma, \gamma) \mathcal{N}(\vect {\beta}_m^{(l)} ; \mu_{\beta}, \sigma^2_{\beta})}{g(\vect{\beta}_m^{(l)} \mid \vect\theta)}}}, 
	\quad \vect{\beta}_m^{(l)}\sim g(\vect{\eta}_m \mid \vect\theta),
	\end{align*}
	where $\widehat P(\vect {y}_{m} \mid \vect {\beta}_m^{(l)}, \sigma, \gamma)$ is the PF estimate of $P(\vect {y}_{m} \mid \vect {\beta}_m^{(l)}, \sigma, \gamma)$.
	
	\subsection{Component-Wise Pseudo-Marginal} \label{sec:CWPM}

	This section defines a component-wise pseudo-marginal (CWPM) method, where the random effects $\vect{{\eta}}_{1:M}$ are updated along with $\vect\theta$. This leads naturally to the following parameter blocks $\vect{\eta}_{1:M}, \{\sigma, \vect{\phi_X}\}$ and $\vect{\phi_\eta}$. If we denote $\vect{\theta_X} = \{\sigma, \vect{\phi_X}\}$, then the joint posterior is of the form
	\begin{align*}
		P(\vect{\theta_X}, \vect{\phi_\eta}, \vect{{\eta}}_{1:M} \mid \vect {y}_{1:M}) \propto P(\vect {y}_{1:M} \mid \vect{{\eta}}_{1:M}, \vect{\theta_X}) P(\vect{{\eta}}_{1:M} \mid \vect{\phi_\eta}) P(\vect{\theta_X}) P(\vect{\phi_\eta}),
	\end{align*}
	and the full conditional posteriors for each of the parameter blocks are
	\begin{eqnarray}
		P(\vect{\eta}_m \mid \vect {y}_{1:M}, \vect{\theta_X}, \phi_\eta) & \propto &
		P(\vect {y}_{1:M} \mid \vect{{\eta}}_{1:M}, \vect{\theta_X}) P(\vect{{\eta}}_{1:M} \mid \vect{\phi_\eta}) \nonumber \\ 
		P(\vect{\theta_X} \mid \vect {y}_{1:M}, \vect{{\eta}}_{1:M}) & \propto & 
		P(\vect {y}_{1:M} \mid \vect{{\eta}}_{1:M}, \vect{\theta_X})P(\vect{\theta_X}) \nonumber \\
		P(\vect{\phi_\eta} \mid \vect{{\eta}}_{1:M}) & \propto &
		P(\vect{{\eta}}_{1:M} \mid \vect{\phi_\eta})P(\vect{\phi_\eta}).
		\label{eq:phiEta}
	\end{eqnarray} 
	 A particle filter estimate of $P(\vect {y}_{1:M} \mid \vect{{\eta}}_{1:M}, \vect{\theta_X})$ is used when updating $\vect{\eta}_{1:M}$ and $\vect{\theta_X}$ (Algorithm \ref{alg:genPF}). The parameter $\vect{\phi_\eta}$ can be sampled directly since \eqref{eq:phiEta} is tractable. This method is generally faster than IAPM as the particle filter is called $2 \times M$ times per MCMC iteration (with the above configuration), instead of $L \times M$ times as in IAPM. If there is a high correlation between $\vect{{\eta}}_{1:M}$ and $\vect\theta$, however, the CWPM chain may mix poorly. 
	
	A correlated version of CWPM (cCWPM) may be implemented using BPM. Again, only the random numbers for a single individual are updated at each iteration while the rest are held constant.
	
	\textbf{Example: SDEMEM with constant drift and diffusion.} For the SDEMEM in Equation \eqref{eq:simpleSDEMEM}, the parameters are updated in the following blocks, $\vect{\eta}_m = \{\beta_m\}$, $\vect{\theta_X} = \{\sigma, \gamma, x_0\}$ and $\vect{\phi_\eta} = \{\mu_\beta, \sigma_\beta\}$.

	\begin{algorithm}[htp]
		initialise $\vect{{\eta}}_{1:M}^{(0)}, \vect{\theta_X}^{(0)}$ and $\vect{\phi_\eta}^{(0)}$ \\
		Run Algorithm \ref{alg:genPF} to obtain the likelihood estimate $\widehat P(\vect {y}_{1:M} \mid \vect{{\eta}}_{1:M}^{(0)}, \vect{\theta_X}^{(0)})$ \\
		\For{$i=1$ to $I$} {
			
			Draw $\vect{{\eta}^*}_{1:M} \sim q(\cdot \mid \vect{{\eta}}_{1:M}^{(i-1)}$ and $\vect {u^*}\sim P(\cdot)$\\
			
			Run Algorithm \ref{alg:genPF} with $\vect {u^*}$ to obtain the likelihood estimate $\widehat P(\vect {y}_{1:M} \mid \vect{{\eta}^{*}}_{1:M}, \vect{\theta_X}^{(i-1)})$ \\
			
			Accept $\vect{{\eta}^*}_{1:M}$ and $\vect {u^*}$ with probability 
			\begin{align*}
			\alpha = \min\left(1, \frac
			{
				\widehat P\left(\vect {y}_{1:M}\mid\vect{\eta^*}_{1:M}, \vect{\theta_X}^{(i-1)}\right)
				P\left(\vect{\eta^*}_{1:M}\mid\vect{\phi_\eta}^{(i-1)}\right)
				q\left(\vect{\eta}_{1:M}^{(i-1)}\mid\vect{\eta^*}_{1:M}\right)
			}{
				\widehat P\left(\vect {y}_{1:M}\mid\vect{\eta}_{1:M}^{(i-1)}, \vect{\theta_X}^{(i-1)}\right)
				P\left(\vect{\eta}_{1:M}^{(i-1)}\mid\vect{\phi_\eta}^{(i-1)}\right)
				q\left(\vect{\eta^*}_{1:M}\mid\vect{\eta}_{1:M}^{(i-1)}\right)
			}\right)
			\end{align*}	
			\vspace{0.3cm} \\
			Draw $\vect{\theta_X^*} \sim q(\cdot \mid \vect{\theta_X}^{(i-1)})$ and $\vect {u^*}\sim P(\cdot)$ \\
			Run Algorithm \ref{alg:genPF} with $\vect {u^*}$ to obtain the likelihood estimate $\widehat P(\vect {y}_{1:M} \mid \vect{{\eta}}_{1:M}^{(i)}, \vect{\theta_X^{*}})$ \\
			
			Accept $\vect{\theta_X^*}$ and $\vect {u^*}$ with probability
			\begin{align*}
				\alpha = \min\left(1,\frac
				{
					\widehat P\left(\vect {y}_{1:M}  \mid \vect{{\eta}}_{1:M}^{(i)}, \vect{\theta_X^{*}}\right) 
					P\left(\vect{\theta_X^{*}}\right) 
					q\left(\vect{\theta_X}^{(i-1)} \mid \vect{\theta_X^{*}}\right)
				}{
					\widehat P\left(\vect {y}_{1:M} \mid \vect{{\eta}}_{1:M}^{(i)}, \vect{\theta_X}^{(i-1)}\right) 
					P\left(\vect{\theta_X}^{(i-1)}\right) 
					q\left(\vect{\theta_X^{*}} \mid \vect{\theta_X}^{(i-1)}\right)
				} \right)
			\end{align*} 
			\vspace{0.3cm}\\
			Draw $\vect{\phi_\eta^*} \sim q(\cdot \mid \vect{\phi_\eta}^{(i-1)})$ \\
			
			Accept $\vect{\phi_\eta^*}$ with probability 
			\begin{align*}
				\alpha = \min\left(1,\frac
				{
					P\left(\vect{{\eta}}_{1:M}^{(i)} \mid \vect{\phi_\eta^*}\right) 
					P\left(\vect{\phi_\eta^*}\right)
					q\left(\vect{\phi_\eta}^{(i-1)} \mid \vect{\phi_\eta^*}\right)
				}
				{
					P\left(\vect{{\eta}}_{1:M}^{(i)} \mid \vect{\phi_\eta}^{(i-1)}\right) 
					P\left(\vect{\phi_\eta}^{(i-1)}\right)
					q\left(\vect{\phi_\eta^*} \mid \vect{\phi_\eta}^{(i-1)}\right)
				} \right)
			\end{align*}
		}	
		\caption{The component-wise pseudo-marginal (CWPM) method.}
		\algorithmfootnote{For the correlated version, new random numbers are only drawn for a single block in steps 4 and 7, not the whole vector.}
		\label{alg:CWPM}
	\end{algorithm}
	
	\subsection{Mixed Particle Method} \label{sec:mixedPM}

	Our final method is a variation of the PMMH $+$ PG algorithm of \citet{gunawan2018}. We use a combination of PMMH and PG to update the parameters $\vect{{\eta}}_{1:M}, \sigma, \vect{\phi_X}$ and $\vect{\phi_\eta}$, depending on the form of the full conditional distributions,
	\begin{eqnarray}
	P(\vect{{\eta}}_{1:M} \mid \vect {y}_{1:M}, \sigma, \vect{\phi_X}, \vect{\phi_\eta}) & \propto & P(\vect {y}_{1:M} \mid \vect{{\eta}}_{1:M}, \sigma, \vect{\phi_X}) P(\vect{{\eta}}_{1:M} \mid \vect{\phi_\eta}) \label{eq:fcETA} \\
	P(\sigma \mid \vect {y}_{1:M}, \vect {x}_{1:M}) & \propto & P(\vect {y}_{1:M} \mid \vect {x}_{1:M}, \sigma) P(\sigma) \label{eq:fcSIGMA} \nonumber \\
	P(\vect{\phi_X} \mid \vect {y}_{1:M}, \vect{{\eta}}_{1:M}, \sigma, \vect{\phi_X}) 
	& \propto & P(\vect {y}_{1:M} \mid \vect{{\eta}}_{1:M}, \sigma, \vect{\phi_X}) P(\vect{\phi_X}) \label{eq:fcPHIX} \\
	P(\vect{\phi_\eta} \mid \vect{{\eta}}_{1:M}) & \propto & P(\vect{{\eta}}_{1:M} \mid \vect{\phi_\eta}) P(\vect{\phi_\eta}).\nonumber
	\label{eq:fcPHIETA}
	\end{eqnarray}
	At each iteration, the invariant path $\vect {x}_{1:M}$ is updated using a conditional particle filter (Algorithm \ref{alg:CSMC}). Where the density $P(\vect {y}_{1:M} \mid \vect{{\eta}}_{1:M}, \sigma, \vect{\phi_X})$ is required, i.e.\ \eqref{eq:fcETA} and \eqref{eq:fcPHIX}, a particle filter estimate is used (PMMH step). Since the full conditionals for $\sigma$ and $\vect{\phi_\eta}$ are tractable, these parameters can be sampled directly. It is important that the likelihood estimate is updated once a new value of $\sigma$ is accepted. This must be done with the same $\vect u$ that was used to estimate the previous likelihood. As with CWPM (Section \ref{sec:CWPM}), mixing of the Markov chain can be poor if high correlation exists between $\vect{{\eta}}_{1:M}$ and $\vect\theta$ and/or $\vect {x}_{1:M}$ and $\sigma$. 
	
	Similarly to IAPM and CWPM, a correlated version of MPM (cMPM) can be implemented using BPM, where $\vect u$ is divided into $M$ blocks based on the individuals $m=1,\ldots,M$. 

	\textbf{Example: SDEMEM with constant drift and diffusion.} For the SDEMEM in Equation \eqref{eq:simpleSDEMEM}, the parameters are updated in the following blocks, $\vect{\eta}_m = \{\beta_m\}$, $\vect{\phi_X} = \{\gamma, x_0\}$, $\vect{\phi_\eta} = \{\mu_\beta, \sigma_\beta\}$ and $\sigma$.
	
	\begin{algorithm}[htp]
		initialise $\vect{{\eta}}_{1:M}^{(0)}, \sigma^{(0)}, \vect{\phi_X}^{(0)}$ and $\vect{\phi_\eta}^{(0)}, \vect {x}_{1:M}^{(0)}$ and $\vect {b}_{1:M}^{(0)}$ \\
		Run Algorithm \ref{alg:genPF} to obtain the likelihood estimate $\widehat P(\vect {y}_{1:M} \mid \vect{{\eta}}_{1:M}^{(0)}, \sigma^{(0)}, \vect{\phi_X}^{(0)})$ \\
		\For{$i=1$ to $I$} {
			
			Draw $\vect{{\eta}^*}_{1:M} \sim q(\cdot \mid \vect{{\eta}}_{1:M}^{(i-1)})$ and $\vect {u^*} \sim P(\cdot)$\\
			
			Run Algorithm \ref{alg:genPF} with $\vect {u^*}$ to obtain the likelihood estimate $\widehat P(\vect {y}_{1:M} \mid \vect{{\eta}^{*}}_{1:M}, \sigma^{(i-1)}, \vect{\phi_X}^{(i-1)})$ \\

			Accept $\vect{{\eta}^*}_{1:M}$ and $\vect {u^*}$ with probability 
			\begin{eqnarray*}
			\alpha = \min\left(1,\frac
			{\widehat P(\vect {y}_{1:M} \mid \vect{{\eta}^{*}}_{1:M}, \sigma^{(i-1)}, \vect{\phi_X}^{(i-1)}) P(\vect{{\eta}^{*}}_{1:M} \mid \vect{\phi_\eta}^{(i-1)}) q(\vect{{\eta}}_{1:M}^{(i-1)} \mid \vect{{\eta}^{*}}_{1:M}}
			{\widehat P(\vect {y}_{1:M} \mid \vect{{\eta}}_{1:M}^{(i-1)}, \sigma^{(i-1)}, \vect{\phi_X}^{(i-1)})P(\vect{{\eta}}_{1:M}^{(i-1)} \mid \vect{\phi_\eta}^{(i-1)}) q(\vect{{\eta}^{*}}_{1:M} \mid \vect{{\eta}}_{1:M}^{(i-1)}} \right)
			\end{eqnarray*}
		
			\vspace{0.3cm}
			Draw $\sigma^* \sim q(\cdot \mid \sigma^{(i-1)})$ \\
			
			Accept $\sigma^*$ with probability 
			\begin{eqnarray*}
			\alpha = \min\left(1,\frac
			{
				P(\vect {y}_{1:M} \mid \vect {x}_{1:M}^{(i-1)}, \sigma^*) P(\sigma^*) q(\sigma^{(i-1)} \mid \sigma^*)
			}
			{
				P(\vect {y}_{1:M} \mid \vect {x}_{1:M}^{(i-1)}, \sigma^{(i-1)}) P(\sigma^{(i-1)}) q(\sigma^* \mid \sigma^{(i-1)})
			} \right)
			\end{eqnarray*}
			
			Run Algorithm \ref{alg:genPF} with $\vect u$ to update $\widehat P(\vect {y}_{1:M} \mid \vect{{\eta}}_{1:M}^{(i)}, \sigma^{(i)}, \vect{\phi_X}^{(i-1)}, \vect u)$ \\
			
			\vspace{0.3cm}
			Draw $\vect{\phi_X^*} \sim q(\cdot \mid \vect{\phi_X}^{(i-1)})$ and $\vect {u^*}\sim P(\cdot)$ \\
			Run Algorithm \ref{alg:genPF} with $u^*$ to obtain the likelihood estimate $\widehat P(\vect {y}_{1:M} \mid \vect{\eta}_{1:M}^{(i)}, \sigma^{(i)}, \vect{\phi_X^{*}})$ \\
			
			Accept $\vect{\phi_X^*}$ and $\vect {u^*}$ with probability 
			\begin{eqnarray*}
				\alpha = \min\left(1,\frac
				{
					\widehat P(\vect {y}_{1:M}  \mid \vect{{\eta}}_{1:M}^{(i)}, \sigma^{(i)}, \vect{\phi_X^{*}}) 
					P(\vect{\phi_X^{*}}) 
					q(\vect{\phi_X}^{(i-1)} \mid \vect{\phi_X^{*}})
				}{
					\widehat P(\vect {y}_{1:M} \mid \vect{{\eta}}_{1:M}^{(i)}, \sigma^{(i)}, \vect{\phi_X}^{(i-1)}) 
					P(\vect{\phi_X}^{(i-1)}) 
					q(\vect{\phi_X^{*}} \mid \vect{\phi_X}^{(i-1)})
				} \right)
			\end{eqnarray*}
			\vspace{0.3cm}
			
			Draw $\vect{\phi_\eta^*} \sim q(\cdot \mid \vect{\phi_\eta}^{(i-1)})$ \\
			
			Accept $\vect{\phi_\eta^*}$ with probability 
			\begin{eqnarray*}
				\alpha = \min\left(1,\frac
				{
					P(\vect{{\eta}}_{1:M}^{(i)} \mid \vect{\phi_\eta^*}) 
					P(\vect{\phi_\eta^*})
					q(\vect{\phi_\eta}^{(i-1)} \mid \vect{\phi_\eta^*})
				}
				{
					P(\vect{{\eta}}_{1:M}^{(i)} \mid \vect{\phi_\eta}^{(i-1)}) 
					P(\vect{\phi_\eta}^{(i-1)})
					q(\vect{\phi_\eta^*} \mid \vect{\phi_\eta}^{(i-1)})
				} \right)
			\end{eqnarray*}
			\vspace{0.3cm}
			
			Run Algorithm \ref{alg:CSMC} with $\vect {x}_{1:M}^{(i-1)}$ and $\vect {b}_{1:M}^{(i-1)}$ to obtain a new path $\vect {x}_{1:M}^{(i)}$ and $\vect {b}_{1:M}^{(i)}$ \\
		}	
	
		\vspace{0.5cm}
		\caption{Mixed particle method (MPM) algorithm. }
		\algorithmfootnote{Random effects $\vect{\eta}_{1:M}$ and parameters $\vect{\phi_X}$ are updated using PMMH. The latent states $\vect {X}_{1:M}$ are updated using PG and $\sigma$ and $\vect{\phi_\eta}$ are updated directly. For the correlated version, new random numbers are only drawn for a single block in steps 4 and 10, not the whole vector.}
		\label{alg:MPM}
	\end{algorithm}

	\subsection{Likelihood Estimation} \label{sec:comp}
		
	We have introduced three particle MCMC methods for SDEMEMs: IAPM, CWPM and MPM. Each of these methods relies on a particle filter to calculate an unbiased estimate of the intractable likelihood. Tuning parameters for this calculation include the level of discretization ($D$), the number of particles ($N$) and, for IAPM, the number of random effects draws ($L$).	We use the log-likelihood ratio $R$ \eqref{eq:ll_ratio} as described in Section \ref{sec:CPM} to tune $D, N$ and $L$. We denote the standard deviation of $|R|$ as $\sigma_\Delta$ and aim for $\sigma_\Delta \le 1.05$.
	
	It is also necessary to specify a proposal function for the particle filter and an importance density for IAPM. Section \ref{sec:SDEsim} describes three different ways to simulate from an SDE: the Euler-Maruyama discretization (EMD), the modified diffusion bridge (MDB) and the residual bridge (RB). Any of these can be used to move particles within a particle filter. Section \ref{sec:IAPM} also proposes the Laplace-ODE and Laplace-MDB importance densities for IAPM. The optimal choice of the proposal function and the importance density is problem specific and may have a large impact on the efficiency of the likelihood estimate.
	
	\section{Example} \label{sec:ex}
	
	\subsection{Data} \label{sec:ex.data}
	
	We apply our methods to real data from a tumour xenography study on mice. This data was obtained from \citet{picchini2019bayesian}. The study had 4 treatment groups and 1 control group, and each group had 7-8 mice. Measurements were taken every Monday, Wednesday and Friday for six weeks; however the majority of the mice were euthanized before the end of the study, once their tumour volumes exceeded $1000$ cubic mm. 
	
\begin{figure}[t]
		\centering
		\includegraphics[height=0.2\textheight,width=0.6\textwidth]{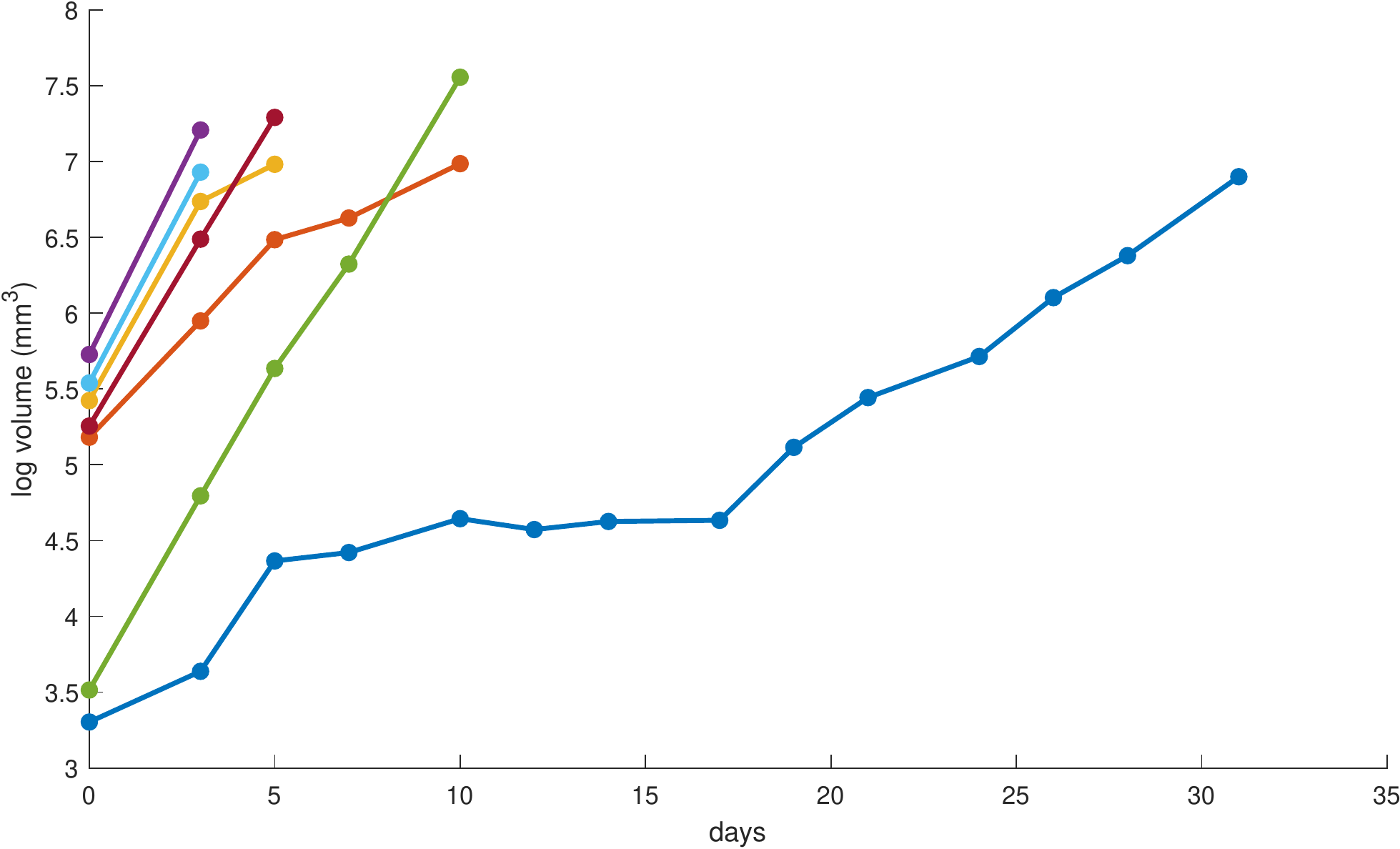}
		\caption{Plot of real tumour volume data.}
		\label{fig:rdata}
	\end{figure}

	We focus specifically on group 5 (the control group). There are 7 mice in this group, with 2-14 observations per mouse and 34 observations in total. Only one mouse in this group survived longer than 11 days, being euthanized on day 32 of the study. Figure \ref{fig:rdata} plots this data. 
	
	\subsection{Model} \label{sec:ex.model}
	To fit the data, we consider an adaptation of an SDEMEM that was used by \citet{picchini2019bayesian} for unperturbed growth. It is assumed that there are $m=1,\ldots,M$ subjects, with measurements taken at discrete times $\xi_t, t=1,\ldots,T_m$, where $T_m$ is the number of observations for subject $m$. The model is defined as, 
	\begin{align}
		dV_{m,t}=\left(\beta_m+\frac{\gamma^2}{2}\right) V_{m,t}dt+\gamma V_{m,t}^\rho dB_{m,t}, \quad V_{m0}=v_{m0},
		\label{eq:untransformed}
	\end{align}
	where $V_{m,t}$ is the volume of subject $m$ at time $t$. The random effects for this model are the parameters $\beta_m$ and $V_{m0}$, which are assigned the prior distributions
	\begin{align*}
		\log(V_{m0}) &\sim \mathcal{N}(\log(V_{m0});\mu_{V0}, \sigma_{V0}^2) \\
		\log(\beta_m) &\sim \mathcal{N}(\log(\beta_m);\mu_{\beta}, \sigma_{\beta}^2).
	\end{align*}
	The observations are modelled as 
	\begin{align}
		Y_{m,t}=\log(V_{m,t}) + \epsilon_{m,t}, \quad \epsilon_{m,t} \sim \mathcal{N}(\epsilon_{m,t};0,\sigma^2).
		\label{eq:utobs}
	\end{align}	
	Since the data is observed on the log scale, the transformation $X_{m,t} = \log(V_{m,t})$ can be applied to \eqref{eq:untransformed} and \eqref{eq:utobs} using It{\^{o}}'s lemma. The full model is then given by
	\begin{equation} 
		\begin{cases}
			Y_{m,t} = X_{m,t} + \epsilon_{m,t}, \quad \epsilon_{m,t} \sim \mathcal{N}(0,\sigma^2) \\
			dX_{m,t} = \left(\beta_m + \frac{\gamma^2}{2}(1-e^{2(\rho-1)X_{m,t}})\right)dt + \gamma e^{(\rho-1)X_{m,t}}dB_{m,t} \\		
			X_{m0} \sim \mathcal{N}(X_{m0};\mu_{X0}, \sigma_{X0}^2) \\
			\log(\beta_m) \sim \mathcal{N}(\log(\beta_m);\mu_{\beta}, \sigma_{\beta}^2).
		\end{cases} 	
		\label{eq:MouseModel}	
	\end{equation}
	The likelihood is intractable since model \eqref{eq:MouseModel} does not have a closed form solution for $X_{m,t}$. The following priors were assigned to the static parameters $\theta = (\mu_{X0}, \sigma_{X0}, \mu_{\beta}, \sigma_{\beta}, \gamma, \sigma, \rho)^T$
	\begin{align*}
	&\mu_{X0} \sim \mathcal{N}(\mu_{X0};3, 4^2) 
	&\sigma_{X0} \sim \mathcal{HN}(\sigma_{X0};5^2) \\
	&\mu_{\beta} \sim \mathcal{N}(\mu_{\beta};0, 4^2) 
	&\sigma_{\beta} \sim \mathcal{HN}(\sigma_{\beta};5^2) \\
	&\gamma \sim \mathcal{HN}(\gamma;5^2) 
	&\sigma \sim \mathcal{HN}(\sigma;5^2) \\
	&\rho \sim \mathcal{N}(\rho;1,0.5^2),
	\end{align*}
	where $\mathcal{HN}(\sigma)$ refers to the half-normal distribution with mean zero and scale parameter $\sigma$.
	
	Note that taking $\rho=1$ gives model \eqref{eq:simpleEx}. This was the original SDEMEM used by \citet{picchini2019bayesian}. We add the parameter $\rho$ which allows for both a more flexible variance and renders the transition density intractable. We test this model on the dataset introduced in Section \ref{sec:ex.data}. To ensure numerical stability when simulating from the SDE, we scaled the observation times by the maximum time observed. In addition to the real data, we also apply our methods to synthetic data simulated from model \eqref{eq:MouseModel} using $\vect\theta = (\mu_{X0}, \sigma_{X0}, \mu_{\beta}, \sigma_{\beta}, \gamma, \sigma, \rho)^T = (3, 1, -1, 1, 1, 0.5, 1)^T$. 
	
\begin{figure}[t]
		\centering
		\includegraphics[height=0.4\textheight,width=1\textwidth]{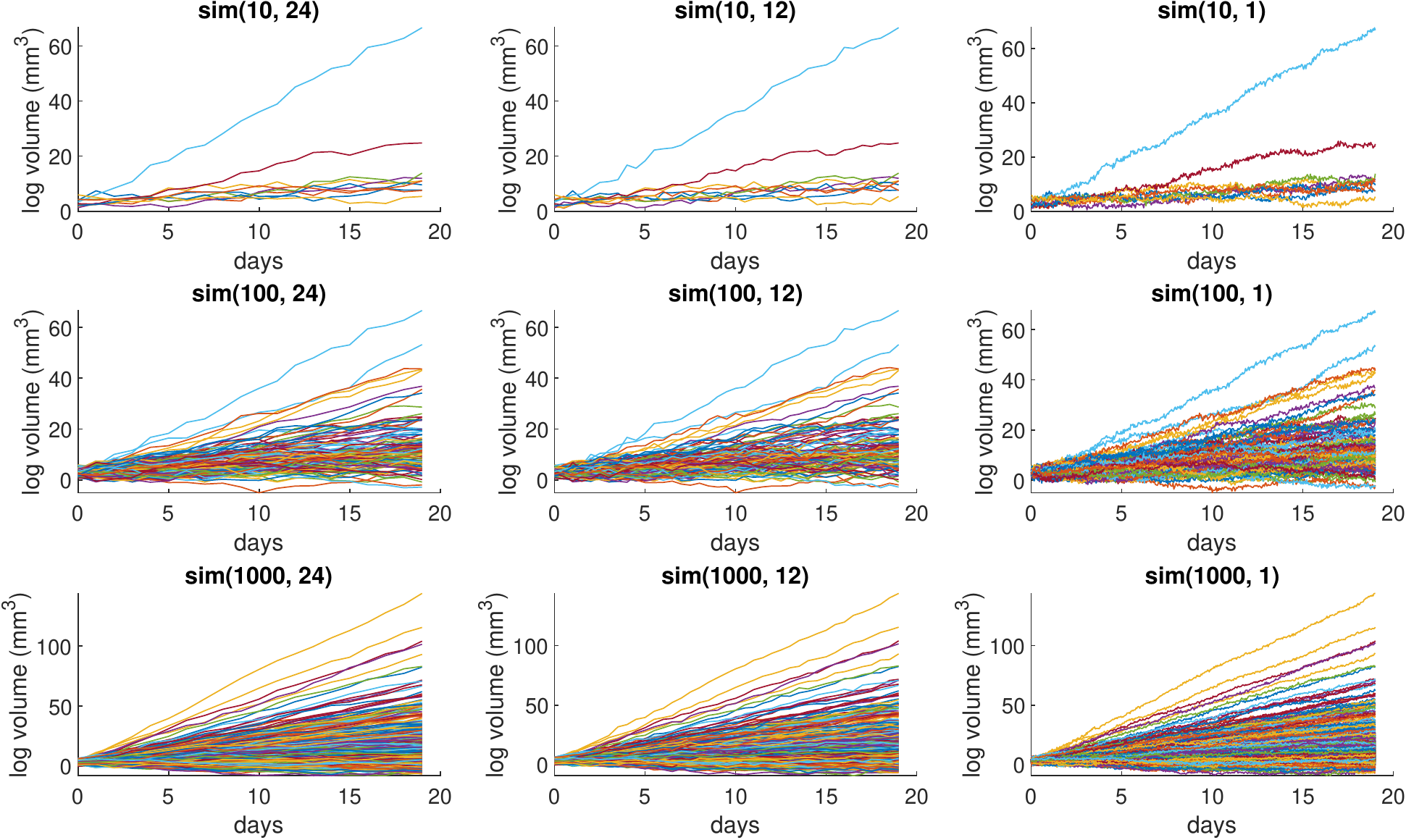}
		\caption{Plot of all simulated datasets. Sim($M,H$) refers to the size of the subset, where $M$ is the number of subjects and $H$ is the number of hours between observations. The full dataset is denoted sim(1000, 1).}
		\label{fig:data}
	\end{figure}

	For the synthetic data, we assumed 1000 mice with 457 observations each - this corresponds to a measurement every hour for 19 days following the initial measurement. We used 9 subsets of this dataset with all combinations of 10, 100 and 1000 subjects and an observation every 24 hours (20 observations), 12 hours (39 observations) and 1 hour (457 observations). We refer to these datasets as sim($M, H$), where $M$ is the number of subjects (10, 100, or 1000) and $H$ is the number of hours between observations (24, 12 or 1). For example, the subset of 100 subjects with an observation every 12 hours is denoted sim(100, 12), while the full dataset is denoted sim(1000, 1). When $M$ is left blank, we refer to all datasets with the specified value of $H$ and vice versa, e.g.\ sim(, 1) represents sim(10, 1), sim(100, 1) and sim(1000, 1). The performance of our methods on these datasets gives an indication of their scalability with respect to the density of the time series and number of subjects. Figure \ref{fig:data} plots this data.

	\section{Likelihood Estimation Results} \label{sec:LE_results}

	All code was implemented in MATLAB. Vectorisation and parallelisation were applied where possible, e.g.\ we used vectorised code for the particle operations and parallelised over the subjects in the particle filter. For IAPM we also parallelised over the random effects draws when running the importance sampler. Our results were calculated using 8 cores. Note that parallelisation was only applied in the particle filter when the average number of observations per subject was greater than 10. We used adaptive resampling in the particle filter when estimating the likelhood. Resampling was done at every iteration in the conditional particle filter. 

	We first consider the efficiency of the likelihood estimation. For each of the three methods, we tested all possible combinations of proposal function and importance density (IAPM). We define the naive method or combination as the IAPM algorithm with the prior as importance density and the Euler-Maruyama approximation as the proposal function in the particle filter. 
		
	As outlined in Section \ref{sec:comp}, we set the tuning parameters such that $\sigma_{\Delta} \le 1.05$. Measurements were calculated from a minimum of $1000$ log-likelihood estimates at a fixed value of $\vect\theta$ and $\vect {\eta}_{1:M}$ (CWPM). For the real data, we used $\vect \theta = (4, 1, 2, 1, 1.6, 0.05, 1)$, which was obtained from a few preliminary MCMC runs (low values of $N, L$ and $D$ were sufficient for this). For the simulated data, we used the true value $\vect \theta = (3, 1, -1, 1, 1, 0.5, 1)$. The random effects $\vect {\eta}_{1:M}$ were determined similarly, using preliminary runs for the real data and the true values for the synthetic data.
	
	We define the level of discretization ($D$) as the number of intermediate timepoints between each observation. We found that the results are not particularly sensitive to this value, so we fixed $D$ at 10 for all methods. Computation was stopped if the computation time for a single log-likelihood estimate exceeded 15 minutes or required more than 150gb of RAM.
	
	In this section, we use the notation `importance density + proposal function' to refer to a particular combination of the two, e.g.\ prior + RB. All combinations were tuned to roughly the same statistical efficiency (based on $\sigma_{\Delta}$), so the most efficient method was taken as the one with the lowest computation time. Further mention of statistical efficiency refers to the value of the tuning parameters $N$ and $L$.
	
	\subsection{IAPM} 
	To tune the IAPM method, we made the simplifying assumption that $N=L$. Tuning was done through trial and error. Of the three methods, IAPM was the most difficult and time-consuming to tune. Assuming $N=L$ simplified the tuning process, but it is not ideal. Depending on the implementation of the code, having a larger/smaller $N$ or $L$ may have a significant impact on the computation time.
	
	Once we started testing combinations, we found that the variance of the Laplace-ODE importance density tends to $0$ for at least one of the random effects, such that the draws for that random effect were close to equal. We solved this by setting the covariance to a diagonal matrix of the prior variances scaled by $0.5$. We denote this altered importance density as L-ODE.
	
	Tables \ref{tab:IAPM_real}-\ref{tab:IAPM_M1000} summarize the log-likelihood results for all datasets. Dashed lines indicate that computation time exceeded the time limit specified in Section \ref{sec:LE_results}. This limit was exceeded for all prior and L-ODE combinations on the sim(1000, ) datasets. For the correlated versions of these, we found that the value of $\sigma_\Delta$ had a very high variance. All versions of IAPM exceeded the time limit on dataset sim(1000, 1). 
	
	For the synthetic data, the Laplace-MDB importance density outperformed the prior and L-ODE in terms of overall efficiency.  Of the latter, the L-ODE showed the poorest performance. Results for the uncorrelated versions are only available for sim(10, 24) and sim(10, 12) and these were also the only datasets with L-ODE combinations that outperformed the prior. Based on these results, the ODE may not a good approximation of the underlying states. A large diffusion coefficient and/or measurement error could account for this. 
	
	The most efficient proposal function depended on the size of the dataset. In terms of statistical efficiency, the MDB and RB have nearly identical results across all datasets, and generally outperforms the EMD. The RB takes slightly longer to run than the MDB however, and both are slower than the EMD. While this had little effect on the smallest datasets, the time difference was significant on the larger ones. The EMD approximation gave the best results on the sim(, 1) datasets. 
	
	Correlating the log-likelihoods generally increased the statistical efficiency. On the larger datasets, this increase is significant, as is the corresponding reduction in computation time. Interestingly, for all sim(10, ) datasets, the uncorrelated Laplace-MDB + EMD was more statistically efficient than the correlated version. The same was also true for Laplace-MDB + MDB and RB on sim(10, 1). 
		
	For the real data, the best results were given by the Laplace-MDB in combination with the MDB or RB. A large gain in statistical efficiency was observed relative to the naive combination, i.e.\ prior + EMD. In the uncorrelated case, the tuning parameters reduced from $L = N = 200$ to $L = N = 4$, and in the correlated from $L=N=90$ to $L=N=3$. A 5.5-fold decrease in time was observed from the uncorrelated naive to the best method. 
	
	\begin{table}[!htp]
		\centering
		\begin{tabular}{|cc||ccc||ccc||ccc|}
			\hline 
			\multicolumn{2}{|c||}{} & \multicolumn{3}{c||}{Prior} & \multicolumn{3}{c||}{L-ODE} & \multicolumn{3}{c|}{Lap-MDB} \\
			PF & Cor. & $L,N$ & $\sigma_{\Delta}$ & time (s) & $L,N$ & $\sigma_{\Delta}$ & time (s) & $L,N$ & $\sigma_{\Delta}$ & time (s) \\
			\hline 
			EMD & No & 200 & 0.99 & 0.21 & 60 & 1.04 & 0.12 & 28 & 0.97 & 0.11 \\
			& Yes & 90 & 1.00 & 0.11 & 30 & 1.02 & 0.10 & 19 & 0.99 & 0.09 \\
			MDB & No & 180 & 1.02 & 0.36 & 35 & 0.87 & 0.11 & 4 & 1.02 & 0.05 \\
			& Yes & 65 & 0.93 & 0.12 & 16 & 0.99 & 0.10 & \bes 3 & \bes 0.94 & \bes 0.04 \\
			RB & No & 180 & 1.02 & 0.38 & 35 & 0.85 & 0.11 & 4 & 1.02 & 0.05 \\
			& Yes & 65 & 0.98 & 0.13 & 16 & 0.98 & 0.11 & \bes 3 & \bes 0.95 & \bes 0.04 \\
			\hline	
		\end{tabular}
		\caption{Log-likelihood results for the IAPM method on the real dataset. The highlighted rows show the combinations which gave the best computation time. }
		\label{tab:IAPM_real}
	\end{table}	
	
	\begin{table}[!htp]
		\centering
		\begin{tabular}{|ccc||ccc||ccc||ccc|}
			\hline
			&&& \multicolumn{3}{c||}{sim(10, 24)} & \multicolumn{3}{c||}{sim(10, 12)} & \multicolumn{3}{c|}{sim(10, 1)} \\
			\hline 
			IS & PF & Cor. & $L,N$ & $\sigma_{\Delta}$ & time (s) & $L,N$ & $\sigma_{\Delta}$ & time (s) & $L,N$ & $\sigma_{\Delta}$ & time (s) \\
			\hline 
			Prior & EMD & No & 250 & 0.97 & 1.69 & 370 & 1.04 & 6.29 & 530 & 0.96 & 134.1 \\
			&  & Yes & 115 & 0.99 & 0.54 & 130 & 1.00 & 1.13 & 335 & 0.99 & 52.58 \\
			& MDB & No & 220 & 0.95 & 3.06 & 220 & 1.03 & 5.84 & 570 & 0.97 & 373.3 \\
			&  & Yes & 95 & 0.91 & 0.86 & 100 & 0.93 & 1.63 & 250 & 1.03 & 80.02 \\
			& RB & No & 220 & 0.96 & 3.07 & 220 & 1.03 & 6.14 & 570 & 0.98 & 385.9 \\
			&  & Yes & 95 & 0.93 & 0.87 & 100 & 0.95 & 1.79 & 250 & 1.01 & 87.62 \\
			\hline
			L-ODE & EMD & No & 220 & 1.04 & 1.31 & 950 & 1.02 & 35.9 & - & - & - \\
			&  & Yes & 60 & 0.99 & 0.32 & 120 & 0.98 & 0.99 & 370 & 1.00 & 62.71 \\
			& MDB & No & 145 & 1.03 & 1.57 & 800 & 1.02 & 55.75 & - & - & - \\
			&  & Yes & 20 & 1.00 & 0.22 & 50 & 1.04 & 0.78 & 310 & 0.98 & 118.0 \\
			& RB & No & 145 & 1.03 & 1.62 & 800 & 1.02 & 60.45 & - & - & - \\
			&  & Yes & 20 &1.00 & 0.21 & 50 & 1.04 & 0.81 & 310 & 0.95 & 126.7 \\
			\hline
			Lap-MDB & EMD & No & 40 & 1.00 & 0.20 & 55 & 0.96 & 0.45 & \bes 150 & \bes 1.03 & \bes 14.22 \\
			&  & Yes & 45 & 1.00 & 0.25 & 75 & 1.02 & 0.57 & 290 & 0.96 & 38.61 \\
			& MDB & No & 8 & 1.01 & 0.12 & 16 & 0.99 & 0.23 & 120 & 0.99 & 26.22 \\
			&  & Yes & \bes 4 & \bes 0.88 & \bes 0.10 & \bes 10 & \bes 0.98 & \bes 0.21 & 190 & 1.00 & 52.18 \\
			& RB & No & 8 & 1.01 & 0.12 & 16 & 0.97 & 0.25 & 120 & 0.97 & 29.70 \\
			&  & Yes & 4 & 0.90 & 0.11 & 10 & 0.97 & 0.25 & 190 & 0.98 & 53.87 \\
			\hline	
		\end{tabular}
		\caption{Log-likelihood results for the IAPM method on the sim(10,) datasets. The highlighted rows show the combinations which give the best computation time. The number of observations for each dataset (from left to right): 200, 390, 4,570.}
		\label{tab:IAPM_M10}
	\end{table}

	\begin{table}[!htp]
		\centering
		\begin{tabular}{|ccc||ccc||ccc||ccc|}
			\hline
			&&& \multicolumn{3}{c||}{sim(100, 24)} & \multicolumn{3}{c||}{sim(100, 12)} & \multicolumn{3}{c|}{sim(100, 1)} \\
			\hline 
			IS & PF & Cor. & $L,N$ & $\sigma_{\Delta}$ & time (s) & $L,N$ & $\sigma_{\Delta}$ & time (s) & $L,N$ & $\sigma_{\Delta}$ & time (s) \\
			\hline 
			Prior & EMD & No & 500 & 1.02 & 52.78 & 500 & 1.01 & 105.7 & - & - & - \\
			&  & Yes & 95 & 1.00 & 3.28 & 110 & 0.96 & 8.3985 & 300 & 1.01 & 419.8 \\
			& MDB & No & 300 & 1.03 & 49.71 & 390 & 1.01 & 154.0 & - & - & - \\
			&  & Yes & 45 & 0.97 & 2.88 & 45 & 1.00 & 6.16 & 200 & 1.03 & 556.8 \\
			& RB & No & 320 & 0.97 & 60.60 & 390 & 1.00 & 174.0 & - & - & - \\
			&  & Yes & 45 & 0.96 & 3.07 & 45 & 0.95 & 6.05 & 200 & 0.98 & 584.8 \\
			\hline
			L-ODE & EMD & No & - & - & - & - & - & - & - & - & - \\
			&  & Yes & 155 & 1.00 & 7.51 & 370 & 1.03 & 76.64 & - & - & - \\
			& MDB & No & - & - & - & - & - & - & - & - & - \\
			&  & Yes & 100 & 0.97 & 8.13 & 230 & 1.00 & 57.27 & - & - & - \\
			& RB & No & - & - & - & - & - & - & - & - & - \\
			&  & Yes & 100 & 1.00 & 8.75 & 230 & 1.02 & 65.51 & - & - & - \\
			\hline
			Lap-MDB & EMD & No & 130 & 1.00 & 5.55 & 140 & 1.03 & 11.70 & - & - & - \\
			&  & Yes & 65 & 1.04 & 2.42 & 80 & 1.00 & 5.26 & \bes 300 & \bes 0.97 & \bes 417.7 \\
			& MDB & No & 30 & 0.96 & 2.12 & 50 & 0.96 & 7.48 & - & - & - \\
			&  & Yes & \bes 4 & \bes 0.91 & \bes 0.63 & \bes 10 & \bes 0.98 & \bes 1.53 & 200 & 0.99 & 532.1 \\
			& RB & No & 30 & 0.99 & 2.53 & 50 & 0.98 & 7.92 &  - & - & - \\
			&  & Yes & 4 & 0.94 & 0.68 & 10 & 0.99 & 1.71 & 200 & 0.95 & 587.6 \\
			\hline	
		\end{tabular}
		\caption{Log-likelihood results for the IAPM method on the sim(100,) datasets. The highlighted rows show the combinations which give the best computation time. The number of observations for each dataset (from left to right): 2,000, 3,900, 45,700.}
		\label{tab:IAPM_M100}
	\end{table}

	\begin{table}[!htp]
		\centering
		\begin{tabular}{|ccc||ccc||ccc||ccc|}
			\hline
			&&& \multicolumn{3}{c||}{sim(1000, 24)} & \multicolumn{3}{c||}{sim(1000, 12)} & \multicolumn{3}{c|}{sim(1000, 1)} \\
			\hline 
			IS & PF & Cor. & $L,N$ & $\sigma_{\Delta}$ & time (s) & $L,N$ & $\sigma_{\Delta}$ & time (s) & $L,N$ & $\sigma_{\Delta}$ & time (s) \\
			\hline 
			Prior & EMD & No & - & - & - & - & - & - & - & - & - \\
			&  & Yes & - & - & - & - & - & - & - & - & -  \\
			& MDB & No & - & - & - & - & - & - & - & - & - \\
			&  & Yes & - & - & - & - & - & - & - & - & - \\
			& RB & No & - & - & - & - & - & - & - & - & - \\
			&  & Yes & - & - & - & - & - & - & - & - & - \\
			\hline
			L-ODE & EMD & No & - & - & - & - & - & - & - & - & - \\
			&  & Yes & - & - & - & - & - & - & - & - & - \\
			& MDB & No & - & - & - & - & - & - & - & - & - \\
			&  & Yes & - & - & - & - & - & - & - & - & - \\
			& RB & No & - & - & - & - & - & - & - & - & - \\
			&  & Yes & - & - & - & - & - & - & - & - & - \\
			\hline
			Lap-MDB & EMD & No & 350 & 1.04 & 285.5 & 400 & 1.05 & 701.2 & - & - & - \\
			&  & Yes & 65 & 0.97 & 24.95 & 80 & 1.03 & 50.76 & - & - & - \\
			& MDB & No & 90 & 1.04 & 71.46 & 145 & 1.02 & 292.2 & - & - & - \\
			&  & Yes & \bes 4 & \bes 0.92 & \bes 6.50 & \bes 10 & \bes 1.01 & \bes 15.65 & - & - & - \\
			& RB & No & 90 & 1.05 & 77.83 & 145 & 0.98 & 315.3 & - & - & -  \\
			&  & Yes & 4 & 0.94 & 7.20 & 10 & 1.00 & 16.63 & - & - & -  \\
			\hline	
		\end{tabular}
		\caption{Log-likelihood results for the IAPM method on the sim(1000,) datasets. The highlighted rows show the combinations which give the best computation time. The number of observations for each dataset (from left to right): 20,000, 39,000, 457,000.} 
		\label{tab:IAPM_M1000}
	\end{table}

	\subsection{CWPM} \label{sec:LE_results_CWPM}
	For CWPM, it was only necessary to select a proposal function and find a value for $N$. Again, this was done through experimentation. Tables \ref{tab:CWPM_real}-\ref{tab:CWPM_M1000} show results for all datasets. Dashed lines indicate that the memory limit specified in Section \ref{sec:LE_results} was exceeded. 
	
	For the synthetic datasets, the correlated version had the best results across all proposal functions. The number of particles needed for the standard version grew quickly with the size of the dataset. Also, since the correlation induced is approximately $1-1/M$, the correlated version showed greater improvement as the number of subjects increased. 

	As with IAPM, the most efficient proposal function depends on the size of the dataset. The best results are given by the MDB/RB, MDB and EMD for the sim(, 24), sim(, 12) and sim(, 1) datasets respectively. For the sim( , 1) datasets, any benefit in statistical efficiency from the bridges was outweighed by the increase in computation time. 
	
	For the real data, we found that a single particle was sufficient to obtain $\sigma_{\Delta} \le 1.05$ when using MDB or RB. 
	
	\begin{table}[!htp]
		\centering
		\begin{tabular}{|cc||ccc|}
			\hline 
			PF & Cor. & $L,N$ & $\sigma_{\Delta}$ & time (s) \\
			\hline 
			EMD & No & 200 & 1.02 & 0.0044 \\
			& Yes & 60 & 0.98 & 0.0030 \\
			MDB & No & \bes 1 & \bes 0.34 & \bes 0.0023 \\
			& Yes & \bes 1 & \bes 0.16 & \bes 0.0023 \\
			RB & No & 1 & 0.33 & 0.0024 \\
			& Yes & 1 & 0.17 & 0.0024 \\
			\hline	
		\end{tabular}
		\caption{Log-likelihood results for the CWPM method on the real data. The highlighted rows show the combinations which give the best time.}
		\label{tab:CWPM_real}
	\end{table}

	\begin{table}[!htp]
		\centering
		\begin{tabular}{|cc||ccc||ccc||ccc|}
			\hline
			&& \multicolumn{3}{c||}{sim(10, 24)} & \multicolumn{3}{c||}{sim(10, 12)} & \multicolumn{3}{c|}{sim(10, 1)} \\
			\hline 
			PF & Cor. & $N$ & $\sigma_{\Delta}$ & time (s) & $N$ & $\sigma_{\Delta}$ & time (s) & $N$ & $\sigma_{\Delta}$ & time (s) \\
			\hline
			EMD & No & 450 & 1.02 & 0.0578 & 700 & 1.02 & 0.0915 & 3100 & 1.03 & 1.8672 \\
			& Yes & 65 & 1.00 & 0.0559 & 85 & 1.02 & 0.0583 & \bes 300 & \bes 0.95 & \bes 0.2773 \\
			MDB & No & 30 & 1.00 & 0.0490 & 110 & 0.96 & 0.0728 & 2100 & 1.02 & 3.1266 \\
			& Yes & \bes 3 & \bes 0.98 & \bes 0.0470 & \bes 10 & \bes 1.04 & \bes 0.0554 & 215 & 1.00 & 0.4922 \\
			RB & No & 30 & 1.00 & 0.0503 & 110 & 0.96 & 0.0743 & 2100 & 1.02 & 3.3687 \\
			& Yes & \bes 3 & \bes 0.97 & \bes 0.0473 & 10 & 1.02 & 0.0578 & 210 & 0.99 & 0.54 \\
			\hline	
		\end{tabular}
		\caption{Log-likelihood results for the CWPM method on the sim(10,) datasets. The highlighted rows show the combinations which give the best time.}
		\label{tab:CWPM_M10}
	\end{table}	
	
	\begin{table}[!htp]
		\centering
		\begin{tabular}{|cc||ccc||ccc||ccc|}
			\hline
			&& \multicolumn{3}{c||}{sim(100, 24)} & \multicolumn{3}{c||}{sim(100, 12)} & \multicolumn{3}{c|}{sim(100, 1)} \\
			\hline 
			PF & Cor. & $N$ & $\sigma_{\Delta}$ & time (s) & $N$ & $\sigma_{\Delta}$ & time (s) & $N$ & $\sigma_{\Delta}$ & time (s) \\
			\hline
			EMD & No & 6500 & 1.04 & 1.1254 & 9000 & 0.99 & 2.9294 & - & - & - \\
			& Yes & 120 & 1.03 & 0.1107 & 120 & 1.05 & 0.1536 & \bes 360 & \bes 1.04 & \bes 2.783 \\
			MDB & No & 350 & 0.97 & 0.2385 & 1200 & 1.00 & 1.0330 & - & - & - \\
			& Yes & \bes 3 & \bes 1.05 & \bes 0.1038 & \bes 11 & \bes 1.03 & \bes 0.1492 & 240 & 0.99 & 3.544 \\
			RB & No & 350 & 0.99 & 0.2527 & 1200 & 0.99 & 1.13 & - & - & - \\
			& Yes & \bes 3 & \bes 1.05 & \bes 0.1035 & 11 & 1.01 & 0.1523 & 240 & 0.99 & 4.019 \\
			\hline	
		\end{tabular}
		\caption{Log-likelihood results for the CWPM method on the sim(100,) datasets. The highlighted rows show the combinations which give the best time.}
		\label{tab:CWPM_M100}
	\end{table}

	\begin{table}[!htp]
		\centering
		\begin{tabular}{|cc||ccc||ccc||ccc|}
			\hline
			&& \multicolumn{3}{c||}{sim(1000, 24)} & \multicolumn{3}{c||}{sim(1000, 12)} & \multicolumn{3}{c|}{sim(1000, 1)} \\
			\hline 
			PF & Cor. & $N$ & $\sigma_{\Delta}$ & time (s) & $N$ & $\sigma_{\Delta}$ & time (s) & $N$ & $\sigma_{\Delta}$ & time (s)  \\
			\hline
			EMD & No & - & - & - & - & - & - & - & - & -  \\
			& Yes & 90 & 1.03 & 0.4702 & 110 & 1.03 & 0.9162 & - & - & - \\
			MDB & No & 3500 & 0.98 & 12.76 & 11000 & 1.04 & 78.72 & - & - & -  \\
			& Yes & \bes 3 & \bes 1.05 & \bes 0.4504 & \bes 12 & \bes 0.99 & \bes 0.8920 & - & - & - \\
			RB & No & 3500 & 1.01 & 13.91 & 11000 & 1.02 & 86.91  & - & - & - \\
			& Yes & 3 & 1.04 & 0.4906 & 12 & 0.96 & 0.9433 & - & - & - \\
			\hline	
		\end{tabular}
		\caption{Log-likelihood results for the CWPM method on the sim(1000,) datasets. The highlighted rows show the combination which give the best time.}
		\label{tab:CWPM_M1000}
	\end{table}	

	\subsection{MPM}
	This method uses the same log-likelihood estimate as CWPM, so no extra tuning was required. When $N>1$, we use the same number of particles for the conditional particle filter as for the standard. When $N=1$, as is the case for the real data (see Section \ref{sec:LE_results_CWPM}), we add an extra particle to account for the invariant path. 
	
	\section{MCMC Results} \label{sec:MCMC}
	We used the time per log-likelihood estimate from Section \ref{sec:LE_results} to determine which methods to run, i.e.\ $\le$ 2 seconds for IAPM, $\le$ 1 second for CWPM and $\le$ 0.5 second for MPM. Each of these was run for 100,000 iterations starting at the same values of $\vect\theta$ that was used in Section \ref{sec:LE_results}. The best proposal function and importance density (for IAPM) from Section \ref{sec:LE_results} was used. Where the MDB and RB proposal functions gave similar results, MDB was the preferred choice. Due to the time constraints, the naive method (uncorrelated IAPM with prior + EMD) was only run on the real and sim(10, 24) datasets. None of the methods were run on the sim(100,1) or sim(1000, 1) datasets. 
	
	We used random walk proposals for the parameters which could not be updated directly, i.e.\ those updated with a PMMH step. In CWPM and MPM, we used the pre-conditioned Metropolis-adjusted Langevin algorithm (MALA) to update the random effects hyperparameters $\{\mu_{X0}, \sigma_{X0}, \mu_{\beta}, \sigma_{\beta}\}$, and in MPM, we used a slice sampler to update $\sigma$. For the proposals, we needed to tune the random walk covariance (also used as the MALA pre-conditioning matrix), and the stepsize for MALA. This was done through experimentation. For CWPM and MPM, we found it was easier to tune the variances for the random effects after a good covariance matrix had been found for $\vect\theta$. 
	
	We compare the methods based on the multivariate effective sample size (multiESS) \citep{vats2015multivariate} of $\vect \theta$ and the computation time in minutes. A score for each method is calculated as the approximate rate of independent samples per minute ($\mathrm{\frac{multiESS}{time}}$). Table \ref{tab:MCMCres} shows the score for each method. Table \ref{tab:ESS_breakdown} shows the breakdown of the multiESS for each update block. Tables \ref{tab:ar_iapm}-\ref{tab:ar_cwpm} (see Appendix A) show the acceptance rates (AR) for the three methods on all datasets and Figure \ref{fig:densities} shows the marginal posteriors of $\vect\theta$ for all datasets. As expected, the marginal posteriors become more precise as the size of the dataset grows (via more subjects and/or more densely observed time series).
	
	A large increase in multiESS was observed between IAPM, and CWPM and MPM on all datasets. This is partly due to the $\vect {X_0}$ hyperparameters. It is clear from Table \ref{tab:ESS_breakdown} that the multiESS for $\{\mu_{X0}, \sigma_{X0}\}$ is always larger than the multiESS for any of the other parameter blocks. Based on this, a more efficient algorithm for this particular example might be to use IAPM for $\vect \theta$ and $\vect \beta$ and CWPM for $\vect {X_0}$. The other reason for the increase in multiESS is the more efficient proposals used for $\vect{\phi_\eta}$ and $\sigma$ (in MPM). For both the real and synthetic data, MPM gave the highest multiESS, followed by CWPM.
		
	Across all datasets, the largest score was given by CWPM. This is due both to the higher multiESS compared to IAPM, and the relatively short computation time. In general, CWPM ran much faster than the other two methods. The exception to this was on the sim(10, 24) dataset, where IAPM had the fastest run time. We also note, that MPM sometimes took longer to run than IAPM on the smaller datasets. The reason for this is how parallelisation was applied. As noted before, parallelisation was only implemented within the particle filter if the average number of observations per subject was greater than 10, i.e.\ only on the sim(,12) and sim(,1) datasets. For IAPM however, the importance sampler was always parallelised. As a result, if $L$ is small enough, e.g. less than the number of available cores, then IAPM would not necessarily take longer to run than CWPM or MPM. This also depends on the number of particles needed for the latter two methods.

	\begin{table}[!htp]
		\centering
		\begin{tabular}{|cc||ccc|}
			\hline 
			Data & Method & MultiESS & time (min) & MultiESS/time \\
			\hline
			\hline
			Real & Naive & 802 & 669 & 1.20 \\
			& IAPM & 733 & 77 & 9.48 \\
			&  CWPM & 1431 & 6 & 242.90 \\
			&  MPM & 1719 & 41 & 41.82 \\
			\hline
			\hline
			sim(10, 24) & Naive & 2439 & 4802 & 0.51 \\
			& IAPM & 1289 & 142 & 9.11 \\
			& CWPM & 3064 & 180 & 17.00 \\
			& MPM & 3371 & 448 & 7.53 \\
			\hline	
			sim(10, 12) & Naive & - & - & - \\
			& IAPM & 1504 & 351 & 4.29 \\
			& CWPM & 3028 & 211 & 14.38 \\
			& MPM & 4503 & 479 & 9.39 \\
			\hline	
			sim(10, 1) & Naive & - & - & - \\
			& IAPM & - & - & - \\
			& CWPM & 3197 & 2127 & 1.50 \\
			& MPM & 5607 & 4786 & 1.17 \\
			\hline	
			\hline
			sim(100, 24) & Naive & - & - & - \\
			& IAPM & 1174 & 971 & 1.21 \\
			& CWPM & 3012 & 430 & 7.01 \\
			& MPM & 3663 & 1088 & 3.37 \\
			\hline	
			sim(100, 12) & Naive & - & - & - \\
			& IAPM & 1181 & 2849 & 0.41 \\
			& CWPM & 2485 & 706 & 3.52 \\
			& MPM & 3541 & 1634 & 2.17 \\
			\hline	
			\hline
			sim(1000, 24) & Naive & - & - & - \\
			& IAPM & - & - & - \\
			& CWPM & 2742 & 1609 & 1.70 \\
			& MPM & 3402 & 4644 & 0.73 \\
			\hline	
			sim(1000, 12) & Naive & - & - & - \\
			& IAPM & - & - & - \\
			& CWPM & 1875 & 3158 & 0.59 \\
			& MPM & - & - & - \\
			\hline
		\end{tabular}
		\caption{MCMC results for all methods on all datasets. Results are calculated from chains of length 100,000. Dashed lines indicate that the method was not computationally feasible on that particular dataset.}
		\label{tab:MCMCres}
	\end{table}

	\begin{table}[!htp]
		\centering
		\begin{tabular}{|cc||c||ccc|}
			\hline 
			Data & Method & $\vect \theta$ & $(\gamma, \sigma, \rho)$ & $(\mu_{X0}, \sigma_{X0})$ & $(\mu_{\beta}, \sigma_{\beta})$ \\
			\hline
			\hline 
			Real & CWPM & 1431.4 & 515.3 & 4520.5 & 1279.3 \\
			&  MPM & 1718.9 & 716.4 & 4313.5 & 1588.2 \\
			\hline
			sim(10, 24) & CWPM & 3063.8 & 1197.6 & 8576.3 & 3452.5 \\
			& MPM & 3370.7 & 1821.9 & 9114.9 & 2530.0 \\
			\hline	
			sim(10, 12) & CWPM & 3027.8 & 1397.9 & 8710.3 & 2931.9 \\
			& MPM & 4503.1 & 3347.2 & 8913.9 & 3113.9 \\
			\hline	
			sim(10, 1) & CWPM & 3197.3 & 1921.2 & 6959 & 2955.3 \\
			& MPM & 5606.9 & 7666.7 & 7502 & 2416.9 \\
			\hline	
			sim(100, 24) & CWPM & 3011.5 & 1026.0 & 8777.7 & 3199.0 \\
			& MPM & 3663.1 & 1451.8 & 10037.0 & 3302.3 \\
			\hline	
			sim(100, 12) & CWPM & 2484.9 & 653.0 & 10660.0 & 3160.1 \\
			& MPM & 3540.7 & 1566.7 & 9439.2 & 2793.2 \\
			\hline	
			sim(1000, 24) & CWPM & 2741.6 & 648.7 & 7014.0 & 3649.6 \\
			& MPM & 3402.4 & 1142.8 & 9989.3 & 3266.7 \\
			\hline	
			sim(1000, 12) & CWPM & 1875.1 & 340.93 & 9180.6 & 2657.2 \\
			& MPM & - & - & - & - \\
			\hline	
		\end{tabular}
		\caption{MultiESS breakdown for each parameter block. The $\vect\theta$ column shows the multiESS for all parameters.}
		\label{tab:ESS_breakdown}
	\end{table}

	\newcommand{\dplot}[1]{\includegraphics[height=0.21\textheight,width=0.45\textwidth]{#1} \vspace{0.2cm}}
	
	\begin{figure}[htp]
		\centering
		\setlength{\tabcolsep}{12pt}
		\scriptsize
		\begin{tabular}{c c}
			\textbf{Real} & \textbf{sim(10, 24)} \\
			\dplot{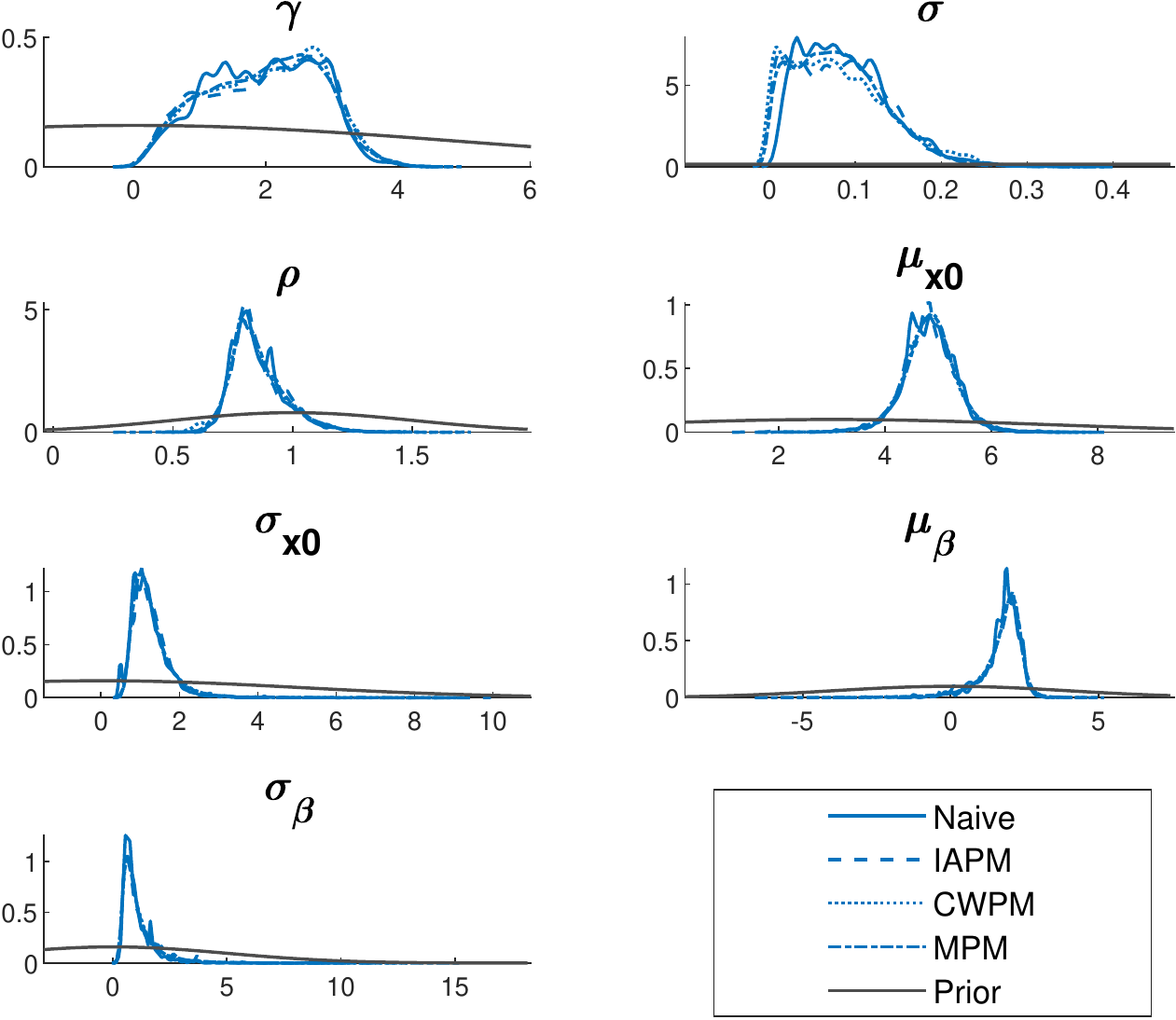} & \dplot{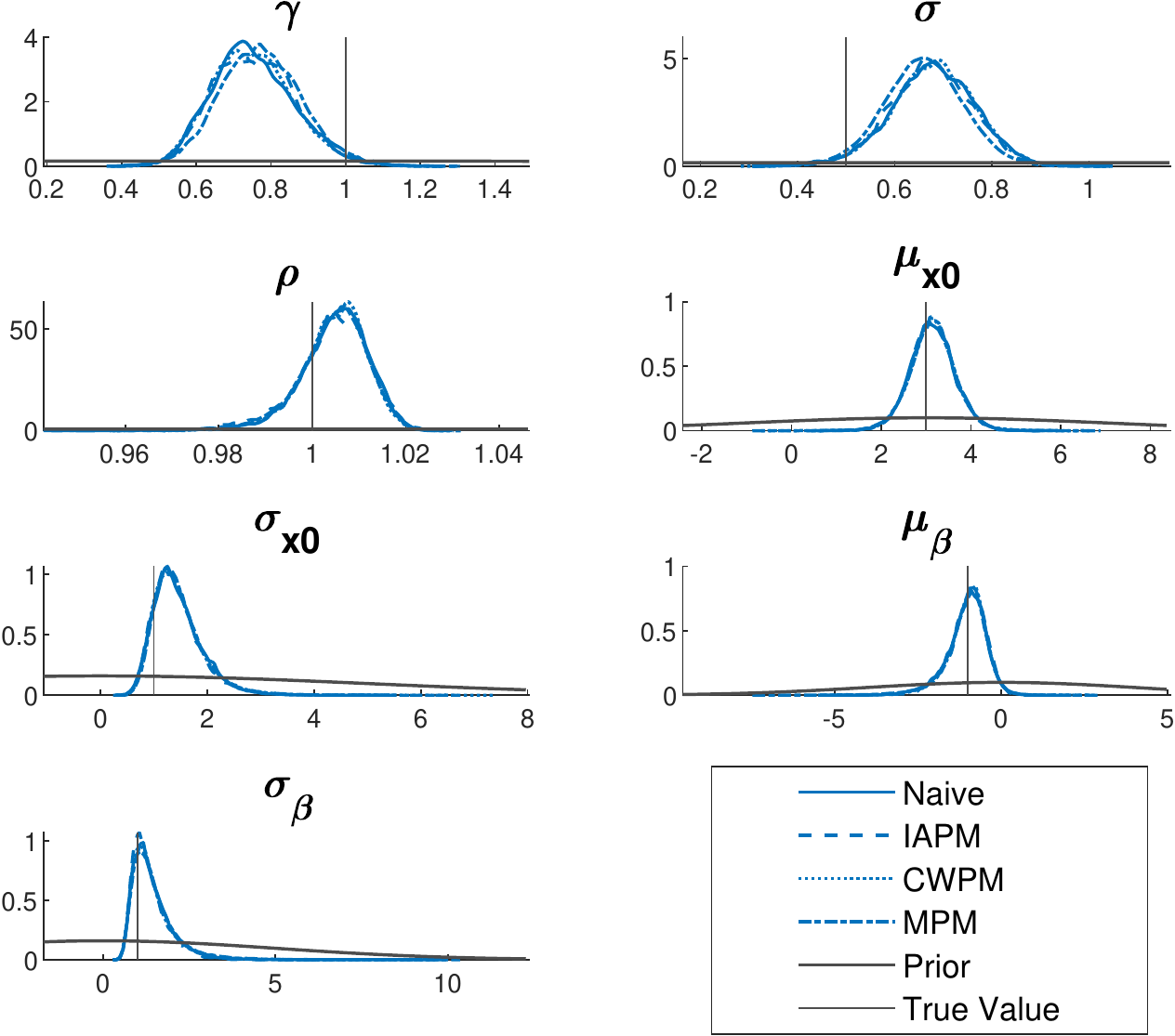} \\ 
			\textbf{sim(10, 12)} & \textbf{sim(10, 1)} \\
			\dplot{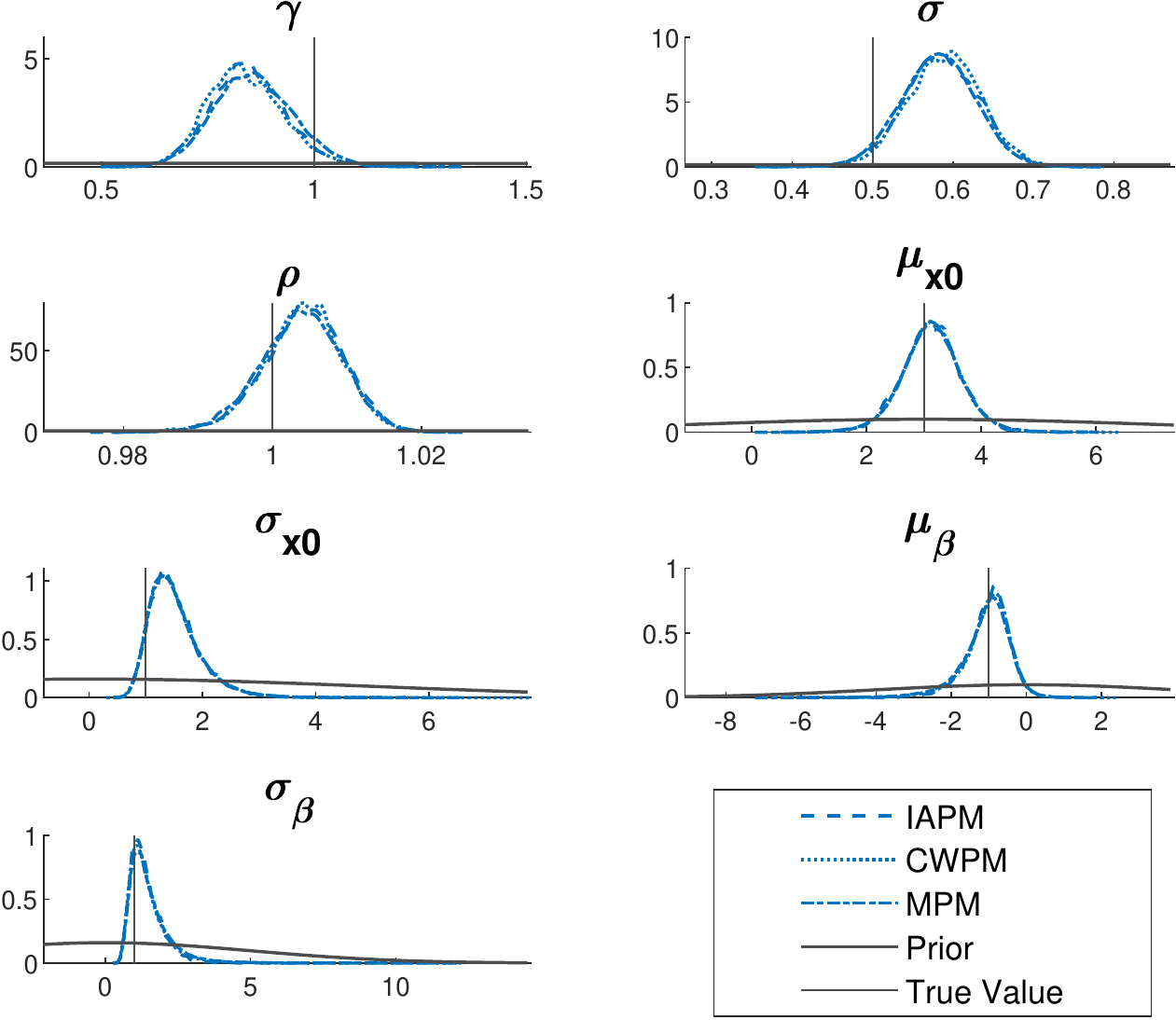} & \dplot{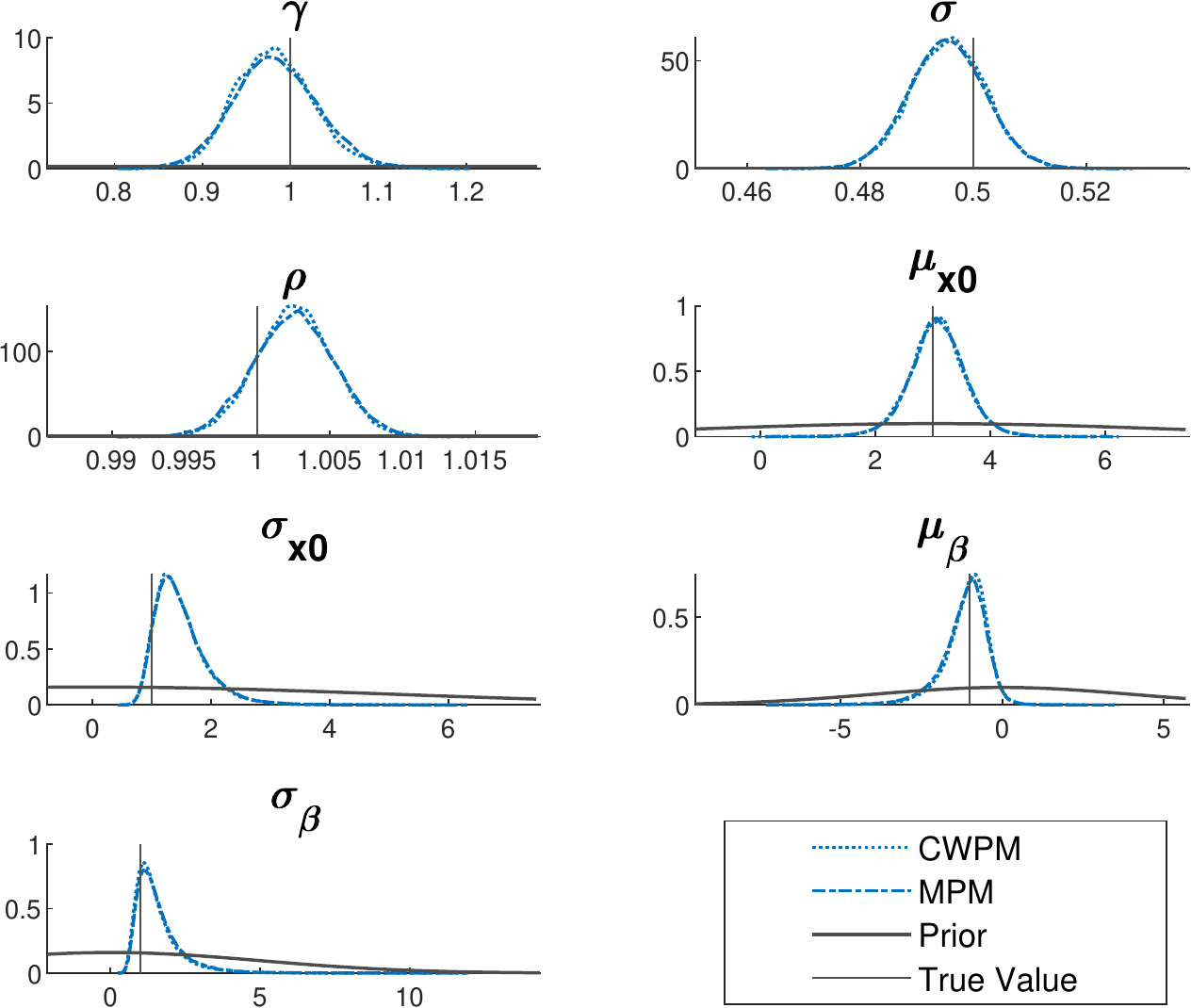} \\ 
			\textbf{sim(100, 24)} & \textbf{sim(100, 12)} \\
			\dplot{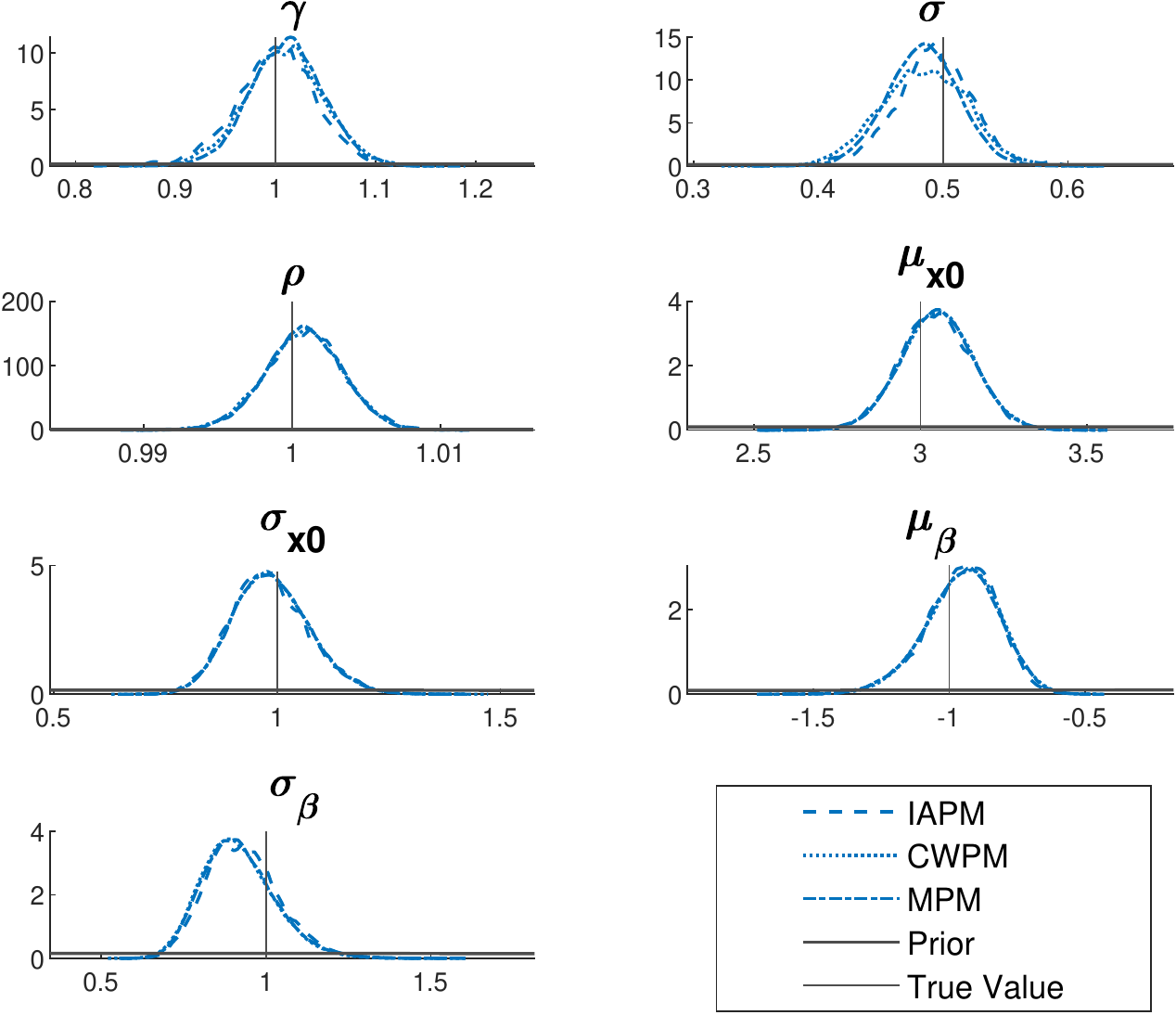} & \dplot{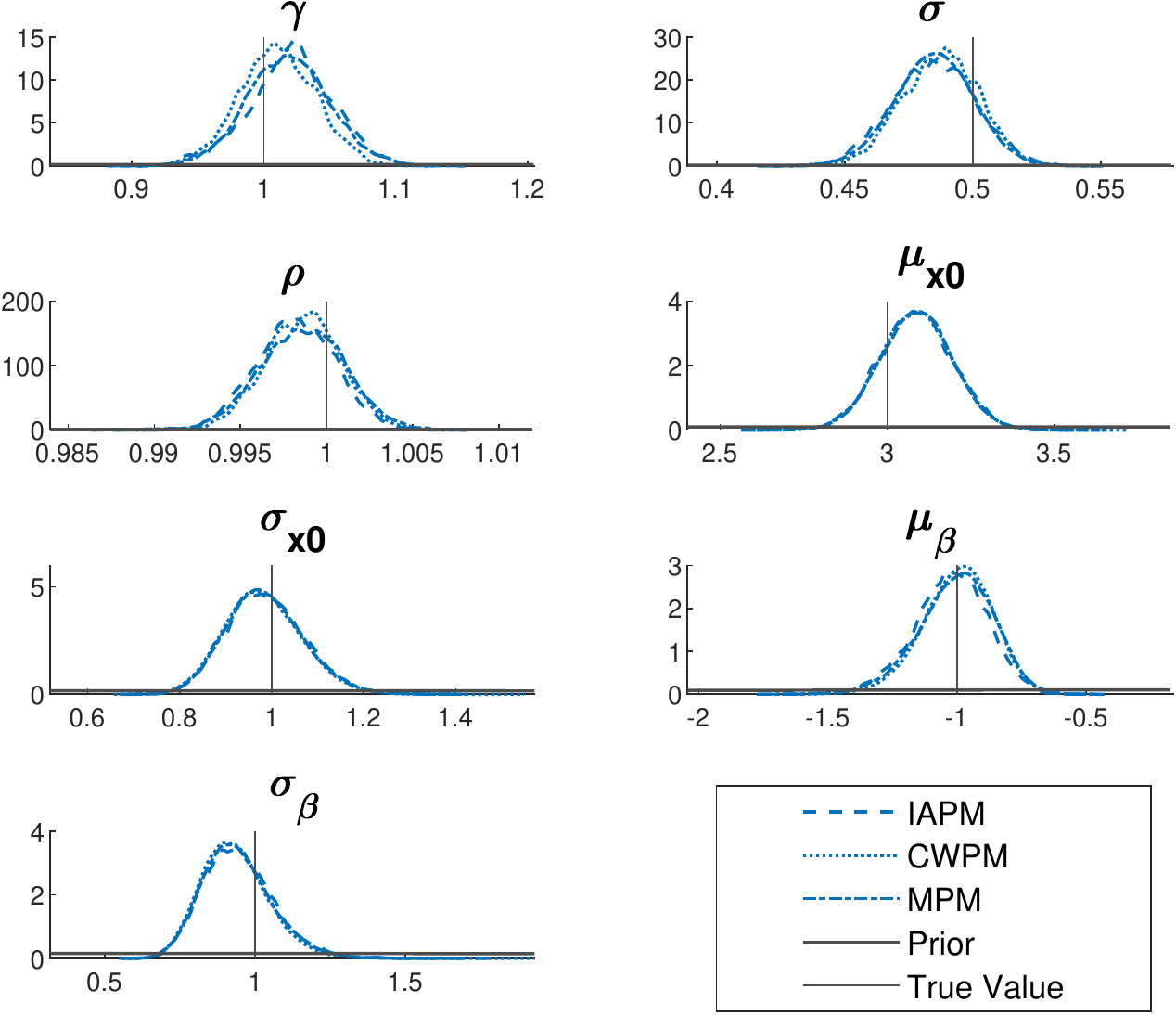} \\ 
			\textbf{sim(1000, 24)} & \textbf{sim(1000, 12)} \\
			\dplot{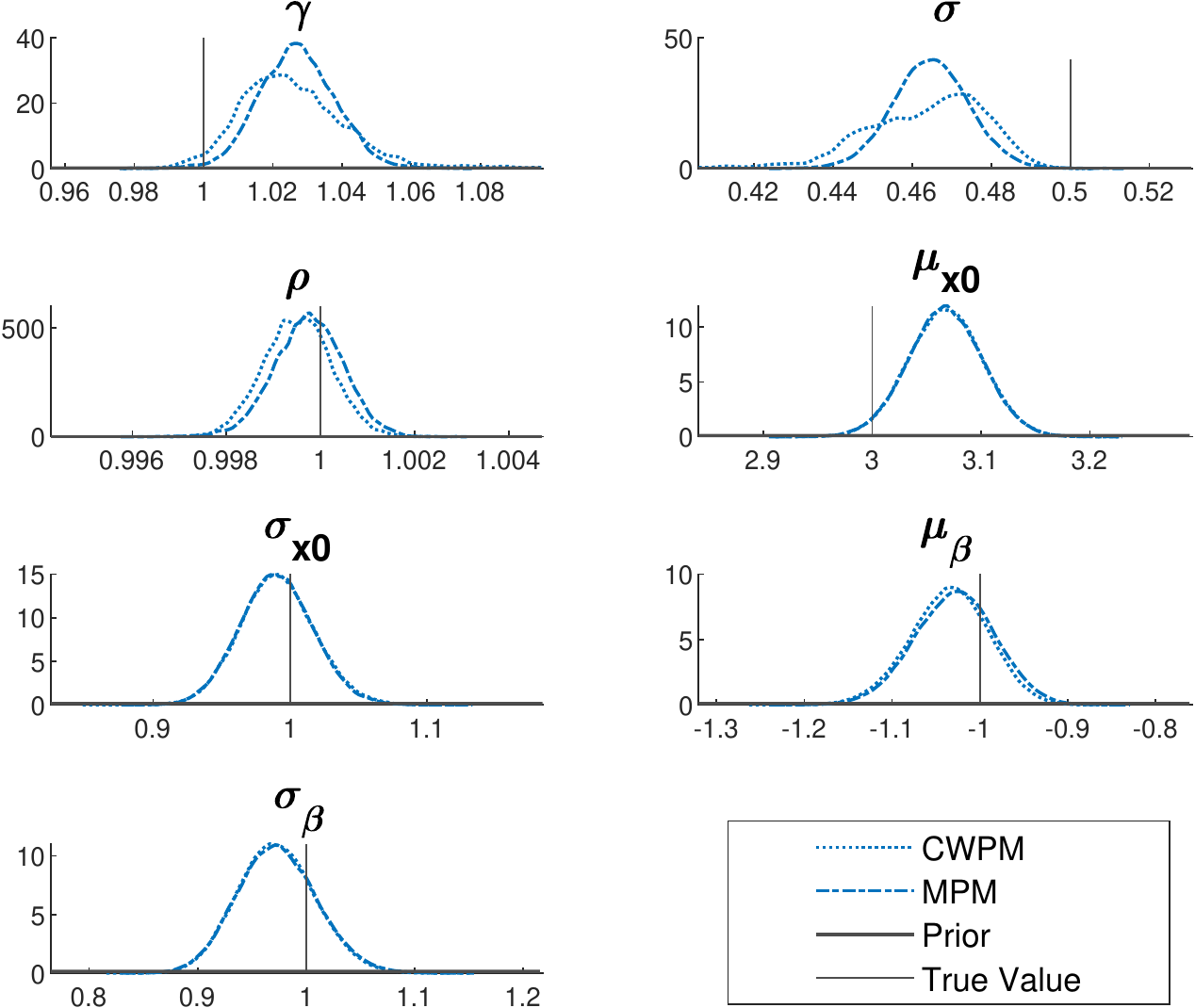} & \dplot{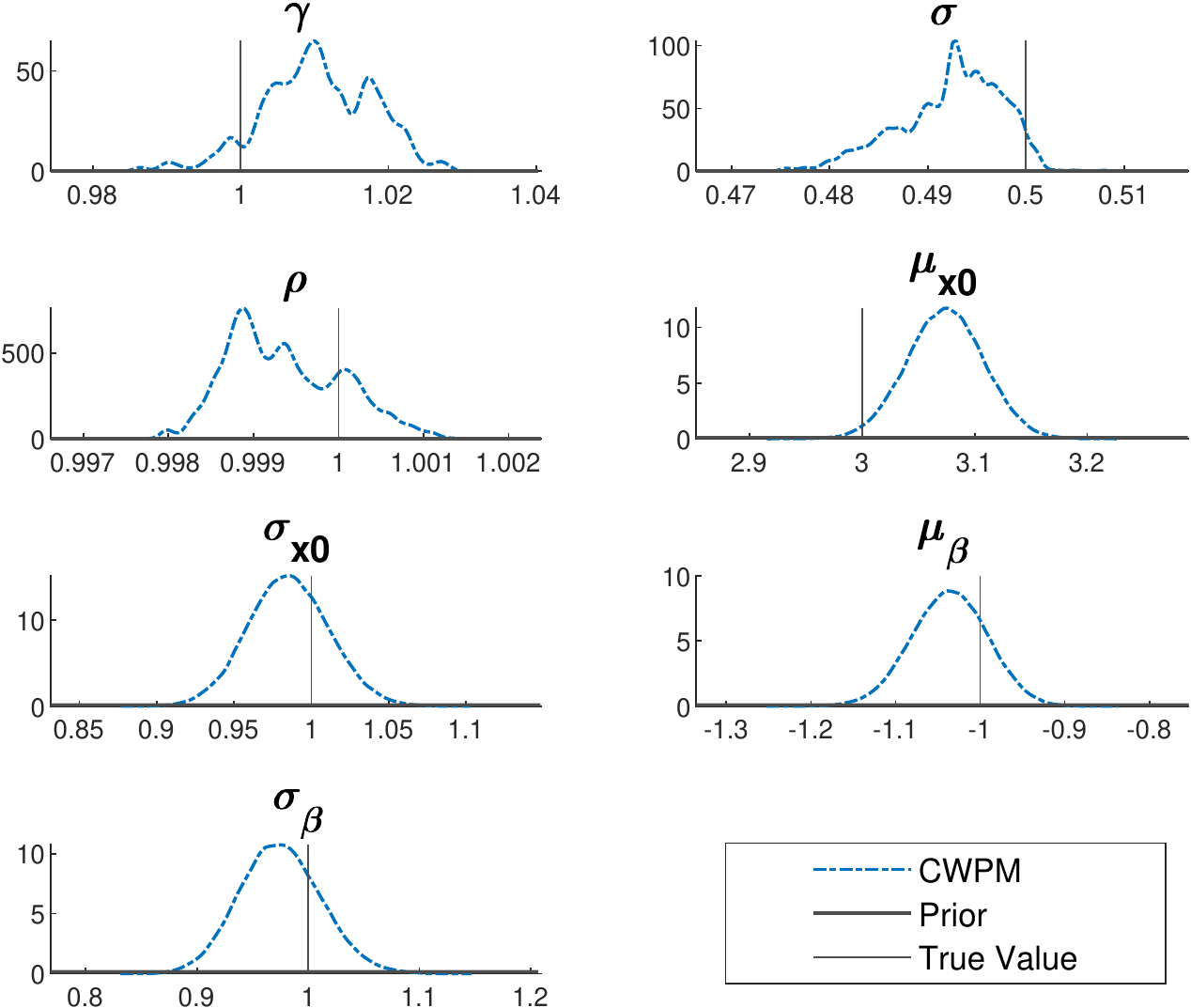} \\ 
		\end{tabular}
		\caption{Univariate posterior density plots of the parameters for all methods and datasets.}
		\label{fig:densities}
	\end{figure}
	
	\section{Discussion} \label{sec:disc}
	
	We introduced three methods for simulation consistent parameter inference of state-space SDEMEMs and outlined some strategies for improving the efficiency of the likelihood estimate for these methods through the choice of importance density and proposal function. The efficiency of the calculation can also be increased by correlating successive log-likelihood estimates. 
	
	The recent paper by \citet[July 23,][]{wiqvist2019particleSDEMEM} independently introduced a method for SDEMEMs that is very similar to our CWPM method. They propose the same update blocks for the parameters as in CWPM and give three variations of this approach, namely naive Gibbs, blocked Gibbs and a correlated PMMH method. In the first, the random numbers $u$ are updated whenever the likelihood is estimated. In blocked Gibbs, $u$ is updated with the random effects but kept fixed for the other parameter blocks. Lastly, their correlated PMMH method uses the approach of \citet{deligiannidis2018correlated} to correlate the likelihoods, i.e.\ by correlating the random numbers (see Section \ref{sec:CPM}). 
	
	Our approach differs in that we use the block pseudo-marginal (BPM) method of \citet{tran2016block}. In the context of mixed effects models, BPM has a number of advantages over CPM. It is simple to implement, induces correlation more directly, and makes no assumptions about the underlying distribution of the random numbers, i.e.\ no transformations to normality are required and it is straighforward to use with RQMC. Also, an efficient implementation only requires the random seed to be stored, which can greatly reduce the computational storage requirements. A drawback of BPM however, is that the correlation is limited by the number of subjects. If there are few subjects, then CPM may be more effective at inducing correlation. Another option might be to combine BPM with CPM, i.e.\ correlating the auxiliary variables in the current block, while keeping the rest fixed. The feasibility of this approach is an area of future research. 
	
	To further improve efficiency, we exploit bridge proposals in the particle filter rather than proposing directly from the (approximate) transition density as in the standard bootstrap filter used by \citet{wiqvist2019particleSDEMEM}. By including the IAPM and MPM methods, our paper provides a more comprehensive suite of particle methods for application to general state-space SDEMEMs. \citet{wiqvist2019particleSDEMEM} allow the number of particles to vary between individuals, which is also straightforward to implement in our methods. 
	
	With IAPM, CWPM and MPM, we were able to greatly improve upon the efficiency of the naive method, particularly in computational efficiency. For the majority of the simulated datasets, the naive approach is not computationally feasible at all. The statistical efficiency of a given method depends on the correlation between the model parameters, random effects and/or latent states. These methods are flexible in the sense that they can be tailored to a  specific model and used in combination, e.g.\ by integrating over a subset of the random effects using IAPM, but updating the rest using CWPM or MPM steps. Note that if IAPM is combined with MPM, then the invariant path from the conditional PF may be used for $\vect{{\widehat{x}}}_m$ in the importance sampler. For our particular example, CWPM gave the best results. In general, this method had the shortest computation time and was the easiest to tune; however as noted before, care must be taken if high correlation exists between the random effects and model parameters. \textit{The best method to use in any particular situation greatly depends on the model and data.}
	
	A significant drawback of all these methods is the amount of tuning required. For all methods (including the naive), there are at least two tuning parameters required for the likelihood estimation. We also do not have a standard way to select the importance density and proposal function, as well as guidelines to indicate whether a correlated pseudo-marginal approach should be used. The last depends on the values of $L$ and $N$, which in turn depend on the efficiency of the method and the dimension of the data. All methods require tuning the MCMC proposal densities. In order to reduce the tuning burden, we plan to embed these methods into a sequential Monte Carlo sampler \citep{del2006sequential} in future research. 
	
	It may be possible to choose the importance density based on the proposal function, i.e.\ EMD + Laplace-ODE (or L-ODE) and MDB/RB + Laplace-MDB. Recall also that the Laplace-ODE approximates the underlying states using the ODE specified by the drift of the SDEMEM. The feasibility of this importance density then relies on how quickly the solution of the ODE can be computed. Exploration of the model could potentially indicate a sensible choice of proposal function and level of discretization $D$. For our example (see Section \ref{sec:ex}), the MDB proposal function generally gives the best results compared to the EMD approximation and RB construct. There are a number of different bridge constructs that can be used however; see \citet{whitaker2017improved} for an overview. The guided proposals of \citet{schauerGuidedProposals} (see also \citet{meulenResidGuided}) are also an option. 	
	
	Lastly, zero-variance control variates \citep{mira2013ZVCV,friel2016exploiting, south2018ZVCV} may be used to further reduce the variance of any expectation estimated from the chains, e.g.\ the expectation of the target with respect to the auxiliary variables. Efficiency of the methods may also be increased through non-centered parameterisations of the random effects $\vect{\eta}_{1:M}$ \citep{omirosNCP2007}.

	\section{Acknowledgments} \label{sec:ack}
	We would like to thank Umberto Picchini and the research team at the Centre for Nanomedicine and Theranostics (DTU Nanotech, Denmark) for providing the real data and Andrew Golightly for useful feedback on an earlier draft of this paper. IB was supported by an Australian Reseach Training Program Stipend and an ACEMS Top-Up Scholarship. IB would also like to thank ACEMS for funding a trip to visit RK at UNSW where some of this research took place. CD was supported by an Australian Research Council's Discovery Early Career Researcher Award funding scheme (DE160100741). The work by RK was partially supported by an ARC Center of Excellence grant (CE140100049). We gratefully acknowledge the computational resources provided by QUT's High Performance Computing and Research Support Group (HPC). 
	
	\bibliographystyle{apalike}
	\bibliography{refs}
	
	\newpage
	
	\section*{Appendix A - Acceptance Rates} 
	
		\begin{table}[!htp]
		\centering
		\begin{tabular}{|c||c|}
			\hline 
			Data & $AR(\vect \theta)$ \\
			\hline
			\hline 
			Real & 0.05 \\
			\hline
			sim(10, 24) & 0.09 \\
			\hline	
			sim(10, 12) & 0.11 \\
			\hline	
			sim(10, 1) & - \\
			\hline	
			sim(100, 24) & 0.10 \\
			\hline	
			sim(100, 12) & 0.11 \\
			\hline	
			sim(1000, 24) & - \\
			\hline	
			sim(1000, 12) & - \\
			\hline	
		\end{tabular}
		\caption{Acceptance rates for the IAPM method based on a chain of length 100,000.}
		\label{tab:ar_iapm}
	\end{table}	
	
	\begin{table}[!htp]
		\centering
		\begin{tabular}{|cc||c|c|}
			\hline 
			Data & Method & $AR(\gamma, \sigma, \rho)$ & $AR(\mu_{X0}, \sigma_{X0}, \mu_{\beta}, \sigma_{\beta})$ \\
			\hline
			\hline 
			Real & CWPM & 0.08 & 0.63 \\
			& MPM & 0.21 & 0.63 \\
			\hline
			sim(10, 24) & CWPM & 0.12 & 0.60 \\
			& MPM & 0.16 & 0.61 \\
			\hline	
			sim(10, 12) & CWPM & 0.12 & 0.56 \\
			& MPM & 0.18 & 0.57 \\
			\hline	
			sim(10, 1) & CWPM & 0.15 & 0.57 \\
			& MPM & 0.27 & 0.58 \\
			\hline	
			sim(100, 24) & CWPM & 0.13 & 0.63 \\
			& MPM & 0.20 & 0.63 \\
			\hline	
			sim(100, 12) & CWPM & 0.06 & 0.61 \\
			& MPM & 0.11 & 0.61 \\
			\hline	
			sim(1000, 24) & CWPM & 0.08 & 0.66 \\
			& MPM & 0.15 & 0.66 \\
			\hline	
			sim(1000, 12) & CWPM & 0.01 & 0.64 \\
			& MPM & - & - \\
			\hline	
		\end{tabular}
		\caption{Acceptance rates for the CWPM and MPM methods based on a chain of length 100,000.}
		\label{tab:ar_cwpm}
	\end{table}

\end{document}